\documentclass[12pt,superscriptaddress,amsmath,amssymb,nofootinbib]{revtex4}
\usepackage{graphicx}
\usepackage{dcolumn}
\usepackage{bm}
\usepackage{amssymb}
\usepackage{amsmath}
\usepackage{epsfig}    
\usepackage{color}
\usepackage{slashed}
\usepackage[hypertex]{hyperref}

\begin{document} 
\title{Effects of custodial symmetry breaking \\in the Georgi-Machacek model at high energies}

\author{Simone Blasi}
\email{simone.blasi.91@gmail.com}
\affiliation{INFN, Sezione di Firenze, and Department of Physics and Astronomy, University of Florence, Via
G. Sansone 1, 50019 Sesto Fiorentino, Italy}

\author{Stefania De Curtis}
\email{decurtis@fi.infn.it}
\affiliation{INFN, Sezione di Firenze, and Department of Physics and Astronomy, University of Florence, Via
G. Sansone 1, 50019 Sesto Fiorentino, Italy}

\author{Kei Yagyu}
\email{yagyu@fi.infn.it}
\affiliation{INFN, Sezione di Firenze, and Department of Physics and Astronomy, University of Florence, Via
G. Sansone 1, 50019 Sesto Fiorentino, Italy}

\begin{abstract}
\noindent
The model proposed by Georgi and Machacek enables the Higgs sector to involve isospin triplet scalar fields while retaining a custodial $SU(2)_V$ 
symmetry in the potential and thus ensuring the electroweak $\rho$ parameter to be one at tree level. 
This custodial symmetry, however, is explicitly broken by loop effects of the $U(1)_Y$ hypercharge gauge interaction. 
In order to make the model consistent at high energies, 
we construct the most general form of the Higgs potential without the custodial symmetry, and then we derive the one-loop $\beta$-functions
for all the model parameters. 
Assuming the $\delta_i$ quantities describing the custodial symmetry breaking to be zero at low energy, 
we find that $|\delta_i|$ are typically smaller than the magnitude of the $U(1)_Y$ gauge coupling and the other running parameters in the potential also
at high energy without spoiling perturbativity and vacuum stability. 
We also clarify that the mass degeneracy among the $SU(2)_V$ 5-plet and 3-plet Higgs bosons is smoothly broken by $\sim 0.1\%$ corrections. 
These results show that the amount of the custodial symmetry breaking is well kept under control up to energies close to the theory cutoff. 

\end{abstract}
\maketitle

\newpage 

\section{Introduction}

The discovered scalar particle with a mass of 125 GeV 
at the LHC Run-I experiment shows properties which are consistent with those of the Standard Model (SM) Higgs boson~\cite{higgs}. 
This experimental fact suggests that the Higgs sector should be constructed by at least one isospin doublet scalar field. 
Due to the still poor experimental accuracy, there are various possibilities for extensions of the Higgs sector, which are predicted in many new physics
scenarios,  from the minimal form assumed in the SM. 
Therefore, the open question is then ``what is the true shape of the Higgs sector?''

One of the most important hints to narrow down the structure of the Higgs sector 
comes from the electroweak $\rho$ parameter, 
which is defined by the ratio of the strength of the charged electroweak current to the neutral one at zero momentum transfer. 
It is well known that its experimental value is quite close to unity, and in fact 
the global fit analysis gives  $\rho^{\text{exp}} = 1.00037\pm 0.00023$~\cite{pdg}. 
On the other hand, the tree level $\rho$ parameter
can be expressed by the ratio of the weak gauge boson masses in an arbitrary Higgs sector, which 
is a sum of contributions from the scalar multiplets $\varphi_i$ with hypercharge $Y_i$, isospin $T_i$ and Vacuum Expectation Value (VEV) $v_i$~\cite{HHG}:
\begin{align}
\rho^{\text{tree}} = \frac{m_W^2}{m_Z^2\cos^2\theta_W} = \frac{\sum_j |v_j|^2[T_j(T_j+1) - Y_j^2]}{2\sum_i|v_i|^2Y_i^2}, \label{rho-tree}
\end{align}
where $\theta_W$ is the weak mixing angle. 
Requiring that, in Eq.~(\ref{rho-tree}), the contribution to the numerator equals that to the denominator for a fixed multiplet $\varphi_i$, we obtain: 
\begin{align}
T_i = \frac{1}{2}\left(\sqrt{1+12Y_i^2} - 1\right). \label{rho2}
\end{align}
The combinations of $T_i$ and $Y_i$ satisfying the above equation are $(T_i,Y_i)$ = $(0,0)$, $(1/2,1/2)$ and $(3,2)$\footnote{The next possibility is 
$(T_i,Y_i)= (25/2,15/2)$, but the introduction of such scalar multiplet breaks the perturbative unitarity due to too large 
gauge couplings for component scalar fields~\cite{Logan-uni}. }. 
Therefore, the introduction of the scalar multiplets with the above assignments
does not change the value of $\rho^{\text{tree}}$ from 1, regardless of the value of their VEVs. 

On the contrary, if we introduce scalar multiplets with $T_i \geq 1$ not satisfying Eq.~(\ref{rho2}), 
$\rho^{\text{tree}}$ can be different from 1. 
In such a case, there are two ways to avoid the constraint from $\rho^{\text{exp}}$, namely, 
(i) tuning the exotic VEVs\footnote{Here, the exotic VEV means that of a scalar multiplet not satisfying Eq.~(\ref{rho2}). } to be quite small, 
or (ii) taking an alignment among the exotic VEVs so as to have a custodial $SU(2)_V$ symmetric  potential. 
The former way is evident, since the contribution to the deviation in $\rho^{\text{tree}}$ from unity is proportional to the squared VEVs as seen in Eq.~(\ref{rho-tree}). 
The latter way gives phenomenologically interesting consequences due to non-negligible exotic VEVs. 
One of the most characteristic consequences is seen in the SM-like Higgs boson ($h$) couplings to the weak gauge bosons $hVV$ ($V=W,Z$), which can be 
larger than the SM prediction~\cite{hVV1,hVV2,hVV3}. Such phenomena cannot be realized in non-minimal Higgs sectors constructed only by singlet and/or doublet scalar fields. 

The model by Georgi and Machacek~\cite{GM1,GM2} (hereafter, simply called GM model), whose Higgs sector is composed of one iso-doublet with $Y=1/2$ and 
two iso-triplets with $Y=1$ and $Y=0$, is the simplest\footnote{This mechanism (ii) can be generalized for models with scalar multiplets 
with $T_i > 1$ as discussed in Ref.~\cite{Logan-gen}.} concrete realization which satisfies $\rho^{\text{tree}}=1$ by the requirement (ii) explained above. 
Basic phenomenological properties of the Higgs bosons in the GM model, e.g., decays and productions 
have been discussed in Refs.~\cite{GVW2,God}. 
After the discovery of the 125 GeV Higgs boson, 
the collider phenomenology of the GM model has been discussed in Refs.~\cite{GM-LHC,CY} at the LHC and in  Ref.~\cite{GM-ILC} at future $e^+e^-$ colliders. 

In the GM model, the two triplet fields can be packaged as an $SU(2)_L\times SU(2)_R$ bi-triplet, and the 
doublet Higgs field forms a bi-doublet by itself. 
As a result, the Higgs potential is invariant under the global $SU(2)_L\times SU(2)_R$ symmetry. 
If we take the VEV of the bi-triplet field to be proportional to the $3\times 3$ unit matrix, 
which corresponds to taking the two triplet VEVs to be the same, 
the $SU(2)_L\times SU(2)_R$ symmetry breaks down to the custodial $SU(2)_V$ symmetry. 

However, it is known that this custodial $SU(2)_V$ symmetry is broken at quantum level due to the $U(1)_Y$ hypercharge gauge boson loop effect~\cite{GVW}. 
In this paper, we quantitatively investigate how this custodial $SU(2)_V$ symmetry is broken at high energies by solving the one-loop 
renormalization group equations (RGEs) for scalar quartic couplings. 
We will show that in order to have consistent $\beta$-functions, we need to start from the most general form of the Higgs potential invariant 
under the $SU(2)_L\times U(1)_Y$ gauge symmetry. 
We then numerically evaluate all the running coupling constants with the initial condition that all the $SU(2)_V$-breaking parameters vanish at low energy.  
We find  that the amount of the custodial symmetry breaking is well kept under control, thus making the custodial symmetric scenario also accessible at high energies. 

This paper is organized as follows. 
In Sec.~\ref{sec:pot}, we present the most general form of the Higgs potential in the GM model. 
We then discuss the relation between the general form and the custodial symmetric one, and define 
the limit to recover the latter at tree level. 
In Sec.~\ref{sec:prob}, we clarify the inconsistency in the derivation of the $\beta$-functions starting from the custodial symmetric form of the potential. 
In Sec.~\ref{sec:num}, we first derive the allowed region in the parameter space by bounds from triviality and vacuum stability as a function of the cutoff scale. 
We then calculate the magnitude of parameters describing the custodial symmetry breaking at high energies. 
We also show the prediction of the mass spectrum for the Higgs bosons at the TeV scale. 
Conclusions are given  in Sec.~\ref{sec:con}. 
In App.~\ref{sec:rel}, we list some useful relations between the $SU(2)_L\times SU(2)_R$ bi-doublet and bi-triplet form of the scalar fields and 
the usual $SU(2)_L$ doublet and triplet ones. 
In App.~\ref{sec:mass}, the mass formulae for all the scalar bosons are given in the general case (but assuming the two triplet VEVs to be the same)
and in the custodial symmetric case. 
In App.~\ref{sec:rge}, the analytic expressions for the one-loop $\beta$-functions for all the parameters of the GM model are presented.

\section{The Most General potential for the GM model \label{sec:pot}}

The scalar sector of the GM model is composed of the complex isospin doublet $\phi$ with $Y=1/2$, 
the complex triplet $\chi$ with $Y=1$ 
and the real triplet $\xi$ with $Y=0$ fields. 
These fields can be expressed by
\begin{align}
\phi = \left(
\begin{array}{c}
\phi^+ \\
\phi^0
\end{array}\right),~
\chi = \left(
\begin{array}{cc}
\frac{\chi^+}{\sqrt{2}} & -\chi^{++}\\
\chi^0 & -\frac{\chi^+}{\sqrt{2}}
\end{array}\right),~
\xi = \left(
\begin{array}{cc}
\frac{\xi^0}{\sqrt{2}} & -\xi^+\\
-\xi^- & -\frac{\xi^0}{\sqrt{2}}
\end{array}\right), \label{par}
\end{align}
where the neutral components are parameterized as  
\begin{align}
\phi^0 =\frac{1}{\sqrt{2}}(\phi_r+v_\phi +i\phi_i),\quad 
\chi^0 =\frac{1}{\sqrt{2}}(\chi_r+i\chi_i^0)+v_\chi,\quad 
\xi^0 = \xi_r+v_\xi,   \label{neut}
\end{align}
with $v_\phi$, $v_\chi$ and $v_\xi$ being the VEVs for $\phi^0$, $\chi^0$ and $\xi^0$, respectively. 
The most general form of the Higgs potential, invariant under the $SU(2)_L\times U(1)_Y$ gauge symmetry, is given by
\begin{align}
V(\phi,\chi,\xi)&=m_\phi^2(\phi^\dagger \phi)+m_\chi^2\text{tr}(\chi^\dagger\chi)+m_\xi^2\text{tr}(\xi^2)\notag\\
&+\mu_1\phi^\dagger \xi\phi +\mu_2 [\phi^T(i\tau_2) \chi^\dagger \phi+\text{h.c.}] +\mu_3\text{tr}(\chi^\dagger \chi \xi)+\lambda (\phi^\dagger \phi)^2 \notag\\
&
+\rho_1[\text{tr}(\chi^\dagger\chi)]^2+\rho_2\text{tr}(\chi^\dagger \chi\chi^\dagger \chi)
+\rho_3\text{tr}(\xi^4)
+\rho_4 \text{tr}(\chi^\dagger\chi)\text{tr}(\xi^2)
+\rho_5\text{tr}(\chi^\dagger \xi)\text{tr}(\xi \chi)\notag\\
&+\sigma_1\text{tr}(\chi^\dagger \chi)\phi^\dagger \phi+\sigma_2 \phi^\dagger \chi\chi^\dagger \phi
+\sigma_3\text{tr}(\xi^2)\phi^\dagger \phi 
+\sigma_4 (\phi^\dagger \chi\xi \phi^c + \text{h.c.}), \label{pot_gen}
\end{align}
where $\phi^c=i\tau_2\phi^*$. 
Although $\mu_2$ and $\sigma_4$ can be complex, we assume them to be real for simplicity.  
In this CP-conserving case, the potential is described by 16 independent real parameters. 
Conventionally, the model with the potential given in Eq.~(\ref{pot_gen}) has not been referred to as the GM model. 
Rather, the GM model has been known as the case where the potential has a global $SU(2)_L\times SU(2)_R$ symmetry. 
In this paper, we will regard the model with Eq.~(\ref{pot_gen}) as the generalized GM model. 

Instead of using the scalar fields given in Eq.~(\ref{par}), let us write the potential with the global $SU(2)_L\times SU(2)_R$ symmetry in terms of 
the $SU(2)_L\times SU(2)_R$ bi-doublet $\Phi$ and the bi-triplet $\Delta$ scalar fields:
\begin{align}
\Phi=\left(
\begin{array}{cc}
\phi^{0*} & \phi^+ \\
-\phi^- & \phi^0
\end{array}\right),\quad 
\Delta=\left(
\begin{array}{ccc}
\chi^{0*} & \xi^+ & \chi^{++} \\
-\chi^- & \xi^0 & \chi^{+} \\
\chi^{--} & -\xi^- & \chi^{0} 
\end{array}\right).  \label{eq:Higgs_matrices}
\end{align}
It takes the following form:
\begin{align}
V(\Phi,\Delta)&=m_\Phi^2\text{tr}(\Phi^\dagger\Phi) + m_\Delta^2\text{tr}(\Delta^\dagger\Delta)\notag\\
&+\lambda_1\text{tr}(\Phi^\dagger\Phi)^2
+\lambda_2[\text{tr}(\Delta^\dagger\Delta)]^2
+\lambda_3\text{tr}[(\Delta^\dagger\Delta)^2]
+\lambda_4\text{tr}(\Phi^\dagger\Phi)\text{tr}(\Delta^\dagger\Delta)\notag\\
&+\lambda_5\text{tr}\left(\Phi^\dagger\frac{\tau^a}{2}\Phi\frac{\tau^b}{2}\right)
\text{tr}(\Delta^\dagger t^a\Delta t^b)\notag\\
&+\bar{\mu}_1\text{tr}\left(\Phi^\dagger \frac{\tau^a}{2}\Phi\frac{\tau^b}{2}\right)(P^\dagger \Delta P)^{ab}
+\bar{\mu}_2\text{tr}\left(\Delta^\dagger t^a\Delta t^b\right)(P^\dagger \Delta P)^{ab}, \label{eq:pot}
\end{align}
where $\tau^a$ and $t^a$ ($a=1$--3) are the $2\times 2$ and $3\times 3$ matrix representations of the $SU(2)$ generators, respectively. 
The matrix $P$ is defined as 
\begin{align}
P=\left(
\begin{array}{ccc}
-1/\sqrt{2} & i/\sqrt{2} & 0 \\
0 & 0 & 1 \\
1/\sqrt{2} & i/\sqrt{2} & 0
\end{array}\right). 
\end{align}
The potential given in Eq.~(\ref{eq:pot}) is described by 9 independent terms\footnote{The custodial symmetric potential does not contain any CP-violating parameters.}. 
Taking the vacuum alignment configuration, i.e. $v_\Delta^{} \equiv v_\chi = v_\xi$, the $SU(2)_L\times SU(2)_R$ symmetry is spontaneously broken down to the 
custodial $SU(2)_V$ symmetry, and the electroweak $\rho$ parameter is predicted to be unity at tree level. 

By using the relations presented in App.~A, 
we find the following correspondence between the parameters defined in Eq.~(\ref{pot_gen}) and those defined in Eq.~(\ref{eq:pot}):
\begin{align}
&m_\phi^2=2m_\Phi^2,~ m_\chi^2=2m_\Delta^2,~ m_\xi^2= m_\Delta^2,~
\mu_1=-\frac{\bar{\mu}_1}{\sqrt{2}},~ \mu_2=-\frac{\bar{\mu}_1}{2},~
\mu_3=6\sqrt{2}\bar{\mu}_2,\notag\\
&\lambda=4\lambda_1,~ 
\rho_1=4\lambda_2+6\lambda_3,~ 
\rho_2=-4\lambda_3,~ 
\rho_3=2(\lambda_2+\lambda_3),~
\rho_4=4\lambda_2,~
\rho_5=4\lambda_3,\notag\\
& \sigma_1 =4\lambda_4-\lambda_5,~ 
 \sigma_2 =2\lambda_5,~ 
 \sigma_3 =2\lambda_4,~
 \sigma_4 =\sqrt{2}\lambda_5. 
\label{rel} 
\end{align}
From the above equations, we can express 7 out of the 16 parameters of the potential in Eq.~(\ref{pot_gen}) (let us choose $m_\xi^2$, $\mu_2$, $\rho_{3,4,5}$ and $\sigma_{4,5}$) 
in terms of the others: 
\begin{align}
&m_\xi^2=\frac{1}{2}m_\chi^2,\quad \mu_2=\frac{1}{\sqrt{2}}\mu_1,\notag \\
&\rho_3 = \frac{1}{2}\rho_1+\frac{1}{4}\rho_2,\quad 
\rho_4=\rho_1+\frac{3}{2}\rho_2,\quad
\rho_5=-\rho_2,\quad
\sigma_3 = \frac{1}{2}\sigma_1+\frac{1}{4}\sigma_2,\quad 
\sigma_4 = \frac{1}{\sqrt{2}}\sigma_2. \label{custodial}
\end{align}
It is convenient to describe the effect of the custodial symmetry breaking in terms of the following quantities $\delta_i$ :
\begin{align}
&\delta_1 \equiv m_\xi^2 -\frac{m_\chi^2}{2},~ \delta_2 \equiv  \mu_2 -\frac{\mu_1}{\sqrt{2}},~
\delta_3 \equiv  \rho_3 -\frac{\rho_1}{2}-\frac{\rho_2}{4},~
\delta_4\equiv  \rho_4 -\rho_1-\frac{3}{2}\rho_2,\notag \\
&\delta_5\equiv  \rho_5 +\rho_2,~
\delta_6 \equiv  \sigma_3 -\frac{\sigma_1}{2}-\frac{\sigma_2}{4},~
\delta_7 \equiv  \sigma_4 -\frac{\sigma_2}{\sqrt{2}}. \label{deltas}
\end{align}
We then define the {\it custodial symmetric limit} by $\delta_i \to 0$, where 
the 16 independent parameters of the general potential are consistently reduced to 9. 

The mass formulae for all the physical Higgs bosons are presented in App.~\ref{sec:mass}
for the general case given in Eq.~(\ref{pot_gen}) with the two triplet VEVs $v_\chi$ and $v_\xi$ to be the same. 
This relation $v_\chi = v_\xi$ is weakly  broken at the TeV scale as we will show in Sec.~\ref{sec:num} as long as we take 
$\delta_i \to 0$ at low energy. 
In App.~\ref{sec:mass}, we also derive the mass formulae in the custodial symmetric case, in which 
all the physical Higgs boson states are classified into the 
$SU(2)_V$ 5-plet $(H_5^{\pm\pm},H_5^\pm,H_5^0)$, 3-plet $(H_3^\pm,H_3^0)$ and two singlets ($H$ and $h$), and the masses of the Higgs boson belonging to 
the same $SU(2)_V$ multiplet are degenerate. 
Thus, there are only 4 independent masses for the Higgs bosons, i.e. the mass of the 5-plet $(m_{H_5}^{})$, that of the 3-plet $(m_{H_3}^{})$, 
and those of the two singlets $m_H^{}$ and $m_h$. 
We will identify $h$ to be the discovered Higgs boson at the LHC with a mass of 125 GeV, i.e., $m_h = 125$ GeV. 

Finally, let us discuss the vacuum stability condition, namely the requirement that the potential does not  
fall down into a negative (infinite) value at any direction of the scalar field space. 
In Ref.~\cite{Logan}, 
the vacuum stability condition has been derived in the custodial symmetric case. 
In the general GM model, there are 5 more independent quartic couplings.
The necessary condition to guarantee the vacuum stability is here derived by assuming two non-vanishing complex fields at once.   
Taking into account all the directions, we obtain the following inequalities:
\begin{align}
\begin{split}
&\lambda \geq 0,~~ \rho_3 \geq 0,~~ \rho_1 + \rho_2 \geq 0,~~ \rho_1 + \frac{\rho_2}{2} \geq 0, \\
& \rho_4 + \frac{\rho_5}{2} +\sqrt{2\rho_3(\rho_1+\rho_2)}\geq 0, \\
& \rho_4  +\sqrt{2\rho_3(\rho_1+\rho_2)}\geq 0, \\
& \rho_4 + 2\sqrt{\rho_3(2\rho_1+\rho_2)}\geq 0, \\
& \rho_4 + \rho_5 + 2\sqrt{\rho_3(2\rho_1+\rho_2)}\geq 0, \\
& \sigma_1 + 2\sqrt{\lambda(\rho_1+\rho_2)}\geq 0, \\
& \sigma_1 + \sigma_2 + 2\sqrt{\lambda(\rho_1+\rho_2)}\geq 0, \\
& \sigma_1 + \frac{\sigma_2}{2} + \sqrt{2\lambda(2\rho_1+\rho_2)}\geq 0, \\
& \sigma_3 +  \sqrt{2\lambda\rho_3}\geq 0. 
\end{split}
\label{stability}
\end{align}

Before closing this section, we briefly review the other parts of the Lagrangian related to the Higgs fields. 
The kinetic Lagrangian is given by 
\begin{align}
\mathcal{L}_{\text{kin}}&=\frac{1}{2}\text{tr}(D_\mu \Phi)^\dagger (D^\mu \Phi)
+\frac{1}{2}\text{tr}(D_\mu \Delta)^\dagger (D^\mu \Delta), \label{lkin}
\end{align}
where the covariant derivatives are expressed as 
\begin{align}
D_\mu \Phi =\partial_\mu\Phi -ig_2\frac{\tau^a}{2}W_\mu^a\Phi + ig_1B_\mu \Phi\frac{\tau^3}{2},\\
D_\mu \Delta =\partial_\mu\Delta -ig_2t^aW_\mu^a\Delta + ig_1B_\mu \Delta t^3. \label{cov}
\end{align}
Eq.~(\ref{lkin}) can also be written in terms of the $\phi$, $\chi$ and $\xi$ fields, as:
\begin{align}
\mathcal{L}_{\text{kin}} 
=|D_\mu \phi|^2+\text{tr}[(D_\mu \chi)^\dagger(D^\mu \chi)]+\frac{1}{2}\text{tr}[(D_\mu \xi)^\dagger(D^\mu \xi)],  \label{kin}
\end{align}
with
\begin{align}
\begin{split}
&D_\mu \phi = \left(\partial_\mu -\frac{i}{2}g_2\tau^a W_\mu^a -\frac{i}{2}g_1 B_\mu\right)\phi, \\
&D_\mu \chi = \partial_\mu\chi-\frac{i}{2}g_2[\tau^a W_\mu^a,\chi] -ig_1 B_\mu\chi,\\
&D_\mu \xi = \partial_\mu\xi -\frac{i}{2}g_2[\tau^a W_\mu^a,\xi]. 
\end{split}
\end{align}
The gauge boson masses are then given by 
\begin{align}
m_W^2 = \frac{g_2^2}{4}(v_\phi^2+4v_\chi^2+4v_\xi^2),\quad 
m_Z^2 = \frac{g_2^2}{4\cos^2\theta_W}(v_\phi^2+8v_\chi^2). 
\end{align}
From Eq.~(\ref{rho-tree}), we can see that, in the custodial symmetric case, i.e., $v_\chi = v_\xi = v_\Delta^{}$, $\rho^{\text{tree}}=1$ is satisfied. 
In this limit, it is convenient to introduce the angle $\beta$ relating to 
the two VEVs $v_\phi$ and $v_\Delta^{}$ by $\tan\beta \equiv v_\phi/(2\sqrt{2}v_\Delta^{})$. 
Also, the SM VEV $v$ is identified by $v^2 = v_\phi^2 + 8v_\Delta^2 = (\sqrt{2}G_F)^{-1} \simeq (246$ GeV)$^2$ with $G_F$ being the Fermi constant. 
The Higgs boson couplings to gauge bosons are obtained from Eq.~(\ref{kin}). As it was already mentioned in the previous section, 
the SM-like Higgs boson couplings to gauge bosons $hVV$ ($V=W,Z$) can be larger than the SM prediction:
\begin{align}
\kappa_V^{} \equiv \frac{g_{hVV}^{\text{GM}}}{g_{hVV}^{\text{SM}}} = \sin\beta \cos\alpha -2\sqrt{\frac{2}{3}}\cos\beta\sin\alpha, \label{kv}
\end{align}
where $g_{hVV}^{\text{GM}}$ $(g_{hVV}^{\text{SM}})$ is the $hVV$ coupling in the GM model (SM), and $\alpha$ is the mixing angle between the CP-even Higgs bosons defined 
in Eq.~(\ref{tan2a}). 
Clearly, $\kappa_V^{}$ can be larger than 1, because of the factor $2\sqrt{2/3}$ in the second term of $\kappa_V^{}$, which comes from the Clebsch-Gordan coefficient of the 
$SU(2)_L$ triplet representation field. 

Finally, the Yukawa Lagrangian is given as follows\footnote{In the GM model, 
there is another possible Yukawa term, written as $\overline{L_L^c}i\tau^2 \chi L_L$, which provides Majorana masses for
the left-handed neutrinos. This is known as the type-II seesaw mechanism~\cite{type2}. 
In our paper, we do not take into account this Yukawa coupling, because it is negligibly small 
as compared to the Yukawa couplings for the doublet Higgs field given in Eq.~(\ref{yuk}). }:
\begin{align}
{\cal L}_Y = -y_t\bar{Q}_L^3 i\tau^2\, \phi^*\, t_R^{} -y_b\bar{Q}_L^3 \,\phi\, b_R^{} -y_\tau\bar{L}_L^3 \,\phi\, \tau_R^{} +\text{h.c.}, \label{yuk}
\end{align}
where we only show the third generation fermion part with $Q_L^3 = (t,b)_L^T$ and $L_L^3 = (\nu_\tau,\tau)_L^T$. 
The fermion masses are obtained as $m_f = y_f v \sin\beta/\sqrt{2}$ $(f=t,b,\tau)$ by taking $\langle \phi^0 \rangle = v \sin\beta/\sqrt{2}$. 

\section{Inconsistency in the $\beta$-function calculation for the custodial symmetric case\label{sec:prob}}

As we already explained in the Introduction, 
we encounter an inconsistency in the calculation of the RGEs, if we start from the Higgs potential defined in Eq.~(\ref{eq:pot}). 
The source of such inconsistency is the $U(1)_Y$ gauge interaction in the kinetic Lagrangian for the Higgs fields,
which explicitly breaks the custodial symmetry at tree level. 
In fact, the kinetic Lagrangian given in Eq.~(\ref{lkin}) is not invariant under the transformations $\Phi \to \Phi U_R^\dagger \, (\Delta \to \Delta U_R^\dagger)$, 
where $U_R$ is the $SU(2)_R$ transformation matrix, due to the generator $\tau^3\,(t^3)$. 
This breaking term affects the scalar potential sector at loop level, i.e.,  
there appear additional operators which break the custodial symmetry and cannot be expressed in terms of $\Phi$ and $\Delta$ defined in Eq.~(\ref{eq:Higgs_matrices}). 
We note that this breaking effect due to the  $U(1)_Y$ gauge interaction is also present in the SM. 
In that case, however, the custodial symmetry emerges accidentally after writing down all the possible renormalizable terms in the potential, so that 
no additional operators can be generated radiatively. 
Therefore, there is no such inconsistency in the SM. 

In order to clarify this problem, let us show as an example,  
the calculation of the one-loop $\beta$-functions for the dimensionless couplings $\rho_1$, $\rho_2$ and $\rho_3$ given in Eq.~(\ref{pot_gen}). 
These can be derived by considering the one-loop vertex function for the $\chi_r^4$ term (denoted as $\hat{\Gamma}_{\chi_r^4}$)
and that for the $\xi_r^4$ term (denoted as $\hat{\Gamma}_{\xi_r^4}$) as follows ($\chi_r$ and $\xi_r$ are introduced in Eq.~(\ref{neut})):
\begin{align}
\hat{\Gamma}_{\chi_r^4} &= \Gamma_{\chi_r^4}^{\text{tree}} + \Gamma_{\chi_r^4}^{\text{1PI}},\quad \hat{\Gamma}_{\xi_r^4} = \Gamma_{\xi_r^4}^{\text{tree}} + \Gamma_{\xi_r^4}^{\text{1PI}}, \label{aaa}
\end{align}
where we have separately indicated the tree level and the one-loop 1-Particle Irreducible (1PI) diagram contributions. 
Let us concentrate on the ${\cal O}(g_1^4)$ terms, so that we do not take into account 
the contribution from the wave function renormalization of the scalar fields which provides ${\cal O}(g_1^2)$ terms in the $\beta$-function. 

The terms arising from the tree level diagrams turn out to be:
\begin{align}
\Gamma_{\chi_r^4}^{\text{tree}}      &= -6(\rho_1 + \rho_2), \quad 
\Gamma_{\xi_r^4}^{\text{tree}}       = -12\rho_3 = -6\rho_1 -3\rho_2 -12\delta_3, 
\end{align}
where we used Eq.~(\ref{deltas}). 
From the one-loop 1PI diagrams, we obtain the following contribution to the ${\cal O}(g_1^4)$ term:
\begin{align}
\Gamma_{\chi_r^4}^{\text{1PI}} & = \frac{1}{16\pi^2}18g_1^4\ln \mu^2 + \cdots, \quad \Gamma_{\xi_r^4}^{\text{1PI}}  = 0 + \cdots,  \notag
\end{align}
where we have displayed only terms proportional to $\ln \mu^2$ with $\mu$ being an arbitrary scale from the dimensional regularization. 
Because the renormalized vertex function must not depend on $\mu$, the following equation should be satisfied
\begin{align}
\frac{d}{d\ln \mu}\hat{\Gamma}_{\chi_r^4} = \frac{d}{d\ln \mu}\hat{\Gamma}_{\xi_r^4} = 0, 
\end{align}
from which we obtain 
\begin{align}
\beta(\rho_1)\big|_{g_1^4} &= -\frac{1}{16\pi^2}6g_1^4 - 4\beta(\delta_3), \quad
\beta(\rho_2)\big|_{g_1^4}  = \frac{1}{16\pi^2}12g_1^4+4\beta(\delta_3),  \label{rho1}
\end{align}
where the $\beta$-function for a parameter $X$ is defined by 
\begin{align}
\beta(X) \equiv \frac{d}{d\ln \mu} X. \label{betaf}
\end{align}
Next, let us consider the $\chi^{++}\chi^{--}\chi_r\chi_r$ and 
$\chi^{++}\chi^-\chi^- \chi_r$ vertices. By following the same steps, we get:  
\begin{align}
\hat{\Gamma}_{\chi^{++}\chi^{--}\chi_r \chi_r} &= -2\rho_1 + \frac{1}{16\pi^2}6 g_1^4 \ln \mu^2+ \cdots, \\
\hat{\Gamma}_{\chi^{++}\chi^-\chi^- \chi_r }  &= -\sqrt{2}\rho_2  + \cdots, 
\end{align}
which give
\begin{align}
\beta(\rho_1)\big|_{g_1^4}  = \frac{1}{16\pi^2} 6 g_1^4,~~~ \beta(\rho_2)\big|_{g_1^4}  = 0. \label{rho1_2}
\end{align}
By comparing Eqs.~(\ref{rho1}) and (\ref{rho1_2}), it is clear that we need a non-vanishing contribution from $\delta_3$, otherwise the $\beta$-functions
for the same coupling obtained by considering different vertices have not the same form. 
In particular, compatibility requires:
\begin{align}
\beta(\delta_3) = -\frac{1}{16\pi^2}3 g_1^4. 
\end{align}
Conversely, $\delta_3$ vanishes in the custodial symmetric potential (together with all the other $\delta$-terms), thus giving rise to 
the mentioned inconsistency in the computation of the $\beta$-functions. 
This issue is not particular of $\rho_1$ and $\rho_2$, but rather it is 
common to all the other couplings in the custodial limit. 
Therefore, in order to obtain a consistent description in terms of the RGEs, we need to introduce the custodial symmetry breaking parameters, or 
in other words, we need to start from the most general potential given in Eq.~(\ref{pot_gen}). 
In App.~\ref{sec:rge}, we present the expressions of the one-loop $\beta$-functions for all the 16 parameters of the general potential, those for the 
three gauge couplings, and those for the top and bottom Yukawa couplings. 

\begin{figure}[t]
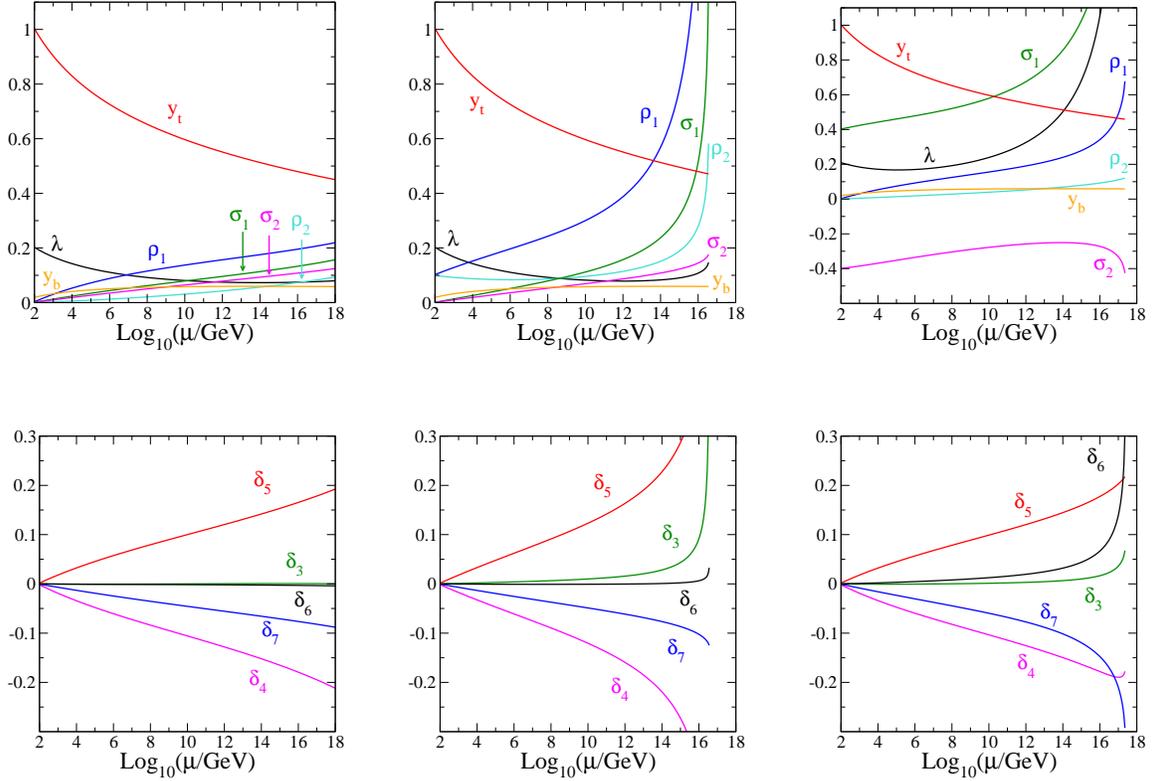

\begin{center}
\includegraphics[width=45mm]{running_0.eps}\hspace{7mm}
\includegraphics[width=45mm]{running_rho1-01.eps}\hspace{7mm}
\includegraphics[width=45mm]{running_sig104_sig2m04.eps}\\\vspace{10mm}
\includegraphics[width=45mm]{running_delta_0.eps}\hspace{7mm}
\includegraphics[width=45mm]{running_delta_rho1-01.eps}\hspace{7mm}
\includegraphics[width=45mm]{running_delta_sig104_sig2m04.eps}
\caption{Running of the dimensionless coupling constants in the case of $\mu_1=100$ GeV, $\mu_3=0$ and $\tan\beta$=5. 
We take the values of ($\rho_1$ $\rho_2$, $\sigma_1$, $\sigma_2$) parameters at the initial scale $\mu = \mu_0(=m_Z^{})$ to be 
(0,0,0,0), (0.1,0.1,0,0) and (0,0,$-0.4$,0.4) for the left, center and right panels, respectively. 
The value of $\lambda$ at $\mu^0$ is fixed to satisfy $m_h=125$ GeV. }
\label{fig1}
\end{center}
\end{figure}

In Fig.~\ref{fig1}, we show the scale dependence of the dimensionless couplings which are evaluated by numerically solving the one-loop RGEs. 
We here take all the $\delta_i$ parameters to be zero at the initial scale $\mu_0 = m_Z^{}$, namely, we assume the custodial symmetric scenario at $\mu_0$. 
The three panels display the running behaviour for three different  configurations of the initial values of the $\rho_1$, $\rho_2$, $\sigma_1$ and $\sigma_2$ parameters. 
We can see that the values of $\delta_i$ become non-zero at  $\mu > \mu_0$ and their magnitudes monotonically increase, but 
the maximal value of $|\delta_i|$ at $\mu > \mu_0$ is typically smaller than the maximal magnitude of the other running scalar couplings at the same scale $\mu$. 
We will further discuss the values of the running $\delta_i$ parameters and their relative size to the other running scalar parameters at high energies in the next section. 
Depending on the initial values, Landau poles can appear at different energy scales, e.g. 
$\mu \sim 10^{16}$ and $\sim 10^{17}$ GeV in the center and right panel of Fig.~\ref{fig1}, respectively. 
Requiring the absence of Landau poles within a certain energy scale constrains the parameter space. 
This feature will be discussed in the next section. 

\section{Numerical results \label{sec:num}}

In this section, we discuss some numerical consequences of the evolution in energy of the couplings of the GM model by using the one-loop RGEs. 
We use the general setup but assuming the custodial $SU(2)_V$ symmetry in the Higgs potential at  low energy
in order to keep the electroweak $\rho$ parameter to be unity. This is realized by taking $\delta_i \to 0$ as defined in Eq.~(\ref{deltas}). 

We first survey the parameter region allowed by the bounds from vacuum stability and triviality as functions of the cutoff scale $\Lambda_{\text{cutoff}}$. 
The former one is defined in such a way that all the inequalities given in Eq.~(\ref{stability})
are satisfied up to $\Lambda_{\text{cutoff}}$, in which all the dimensionless parameters should be understood as functions of the scale $\mu$. 
The latter is defined by requiring that there is no Landau pole up to $\Lambda_{\text{cutoff}}$. 
Here, we impose the following criteria as the triviality bound for all the dimensionless parameters:
\begin{align}
|\lambda(\mu)| \leq 4\pi,~~|\rho_i(\mu)| \leq 4\pi,~~|\sigma_j(\mu)| \leq 4\pi~~~\text{for}~~ \mu_0 \leq \mu \leq \Lambda_{\text{cutoff}}, 
\end{align}
where $i=1,\dots ,5$ and $j=1,\dots ,4$. The initial scale $\mu_0$ is fixed to be $m_Z^{}$. 
In addition to the vacuum stability and triviality bounds, we also require that all the squared masses for the physical Higgs bosons are positive at $\mu_0$. 

We want to show the behaviour of the custodial symmetry breaking parameters $\delta_i$ at high energies according to the evolution of the parameters as given by the RGEs. 
In particular, we want to check if the custodial symmetry is only weakly broken at high energies. 
Since we take the custodial symmetric scenario ($\delta_i \to 0$) at $\mu_0$, all the other parameters at $\mu_0$
are determined according to Eq.~(\ref{custodial}). 

In the numerical analysis, we choose the following 7 parameters in the potential, with $\delta_i = 0$, as inputs: 
\begin{align}
\rho_1^0,~\rho_2^0,~\sigma_1^0,~\sigma_2^0,~\mu_1^0,~\mu_3^0,~\tan\beta^0,
\end{align}
where $X^0 \equiv X(\mu_0)$. Notice that the tadpole conditions  vary by changing $\mu$, so that the value of $\tan\beta$ also depend on $\mu$. For this reason, 
we introduce $\tan\beta^0  =  \tan\beta(\mu_0)$. 
The value of $\lambda^0$ is determined so as to satisfy $m_h=125$ GeV. 

\begin{figure}[t]
\begin{center}
\includegraphics[width=80mm]{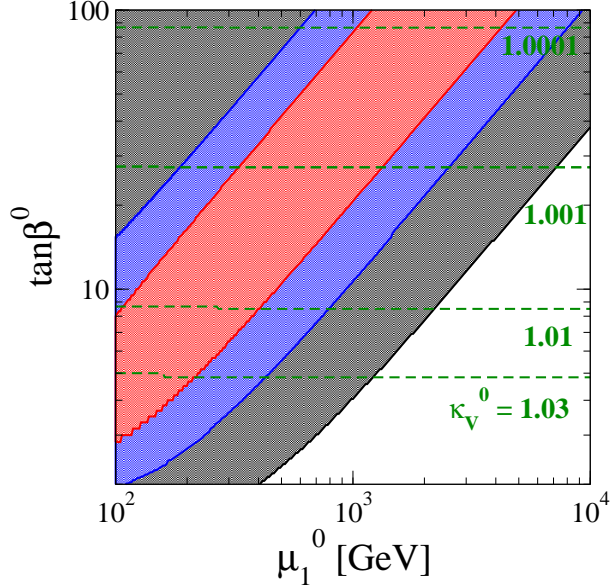}
\caption{
The shaded region is allowed by the triviality and the vacuum stability bounds with the required cutoff scale to be larger than $10^{15}$ GeV (red), 
$10^{8}$ GeV (blue) and $10^{4}$ GeV (black). 
We take $\rho_1^0=\rho_2^0=\sigma_1^0=\sigma_2^0=\mu_3^0=0$.
The green dashed lines show the contour of $\kappa_V^{0}$. }
\label{fig:const1}
\end{center}
\end{figure}

We first consider the case with $\rho_1^0=\rho_2^0 = \sigma_1^0=\sigma_2^0=0$ as a starting point. 
In Fig.~\ref{fig:const1}, we show the allowed parameter space on the $\mu_1^0$--$\tan\beta^0$ plane with $\mu_3^0=0$. 
The black, blue and red shaded regions are allowed from the requirement of $\Lambda_{\text{cutoff}}\geq 10^4$, $10^8$ and $10^{15}$ GeV, respectively. 
In this figure, we also show the contour of the scaling factor $\kappa_V^{0}$ whose tree level formula is given in Eq.~(\ref{kv}). 
We see that the large $\Lambda_{\text{cutoff}}$ is allowed in a limited interval of $\tan\beta^0$ depending on the value of $\mu_1^0$. 
For example, the allowed region with $\Lambda_{\text{cutoff}}\geq 10^{15}$ GeV is obtained in the case with $3\lesssim \tan\beta^0 \lesssim 10$ ($20\lesssim \tan\beta^0 \lesssim 80$)
for $\mu_1^0 = 100$ (1000) GeV. 
This can be understood by the fact that this region requires a smaller value of $\lambda^0$ to satisfy $m_h=125$ GeV as compared to the outside region, which makes 
the appearance of the Landau pole at a higher energy scale.  
We also see that in this configuration, $\kappa_V^{0} > 1$ is predicted in the most of the parameter region on this $\mu_1^0$--$\tan\beta^0$ plane. 
Finally, we checked that the allowed region from the triviality and the vacuum stability bounds 
and the behavior of $\kappa_V^{0}$ do not depend so much on the value of $\mu_3^0$ as long as we take $\mu_3^0$ to be not too large to 
give a negative value of $m_{H_5}^2$. 
In fact, by requiring $m_{H_5}^2 > 0 $, from Eq.~(\ref{m5sq}) we obtain 
\begin{align}
\mu_3^0 < 2\mu_1^0\tan^2\beta^0  + v(\rho_2^0\cos\beta^0  + 3\sigma_2^0 \tan\beta^0 \sin\beta^0 ), 
\end{align}

\begin{figure}[t]
\begin{center}
\includegraphics[width=90mm]{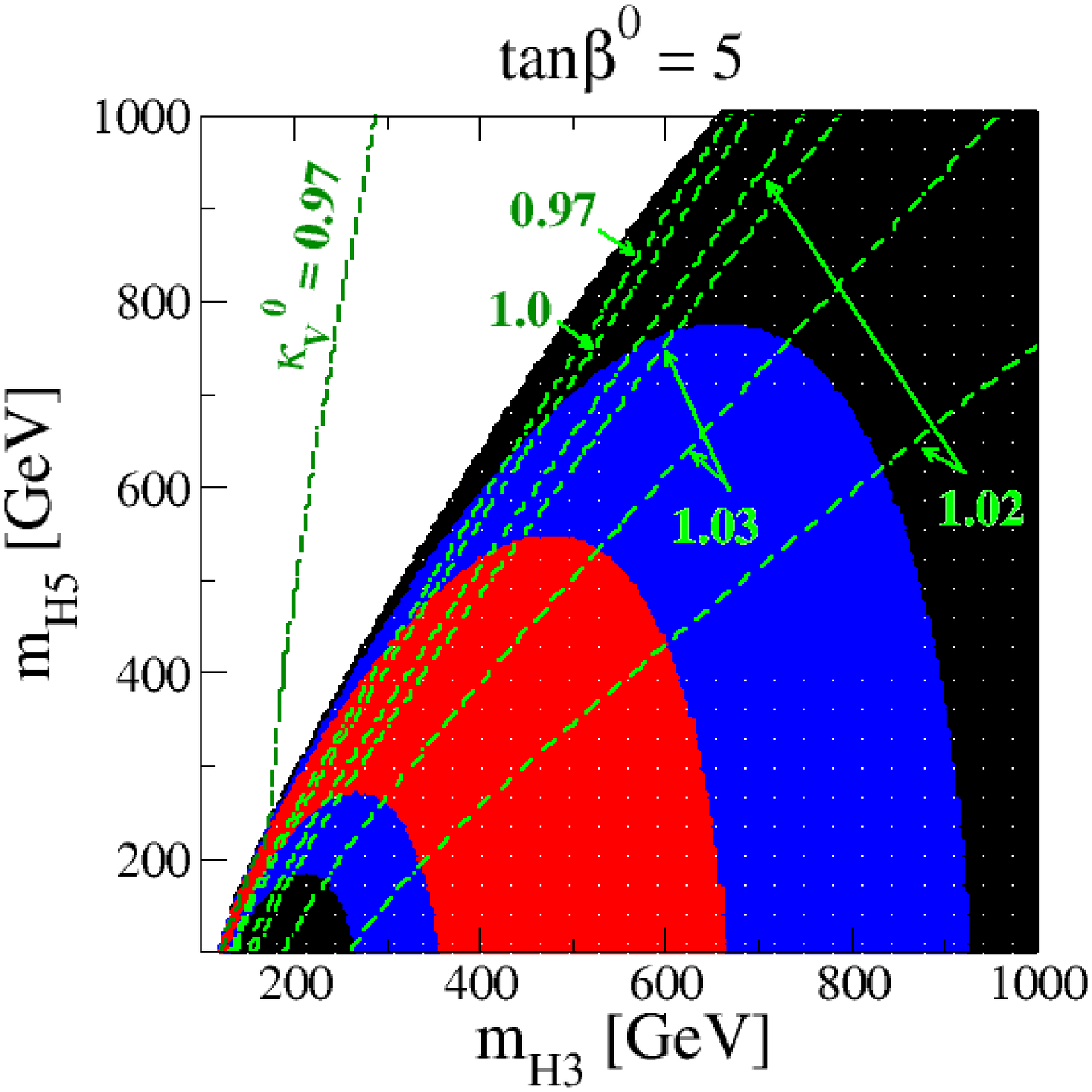}\hspace{-25mm}
\includegraphics[width=90mm]{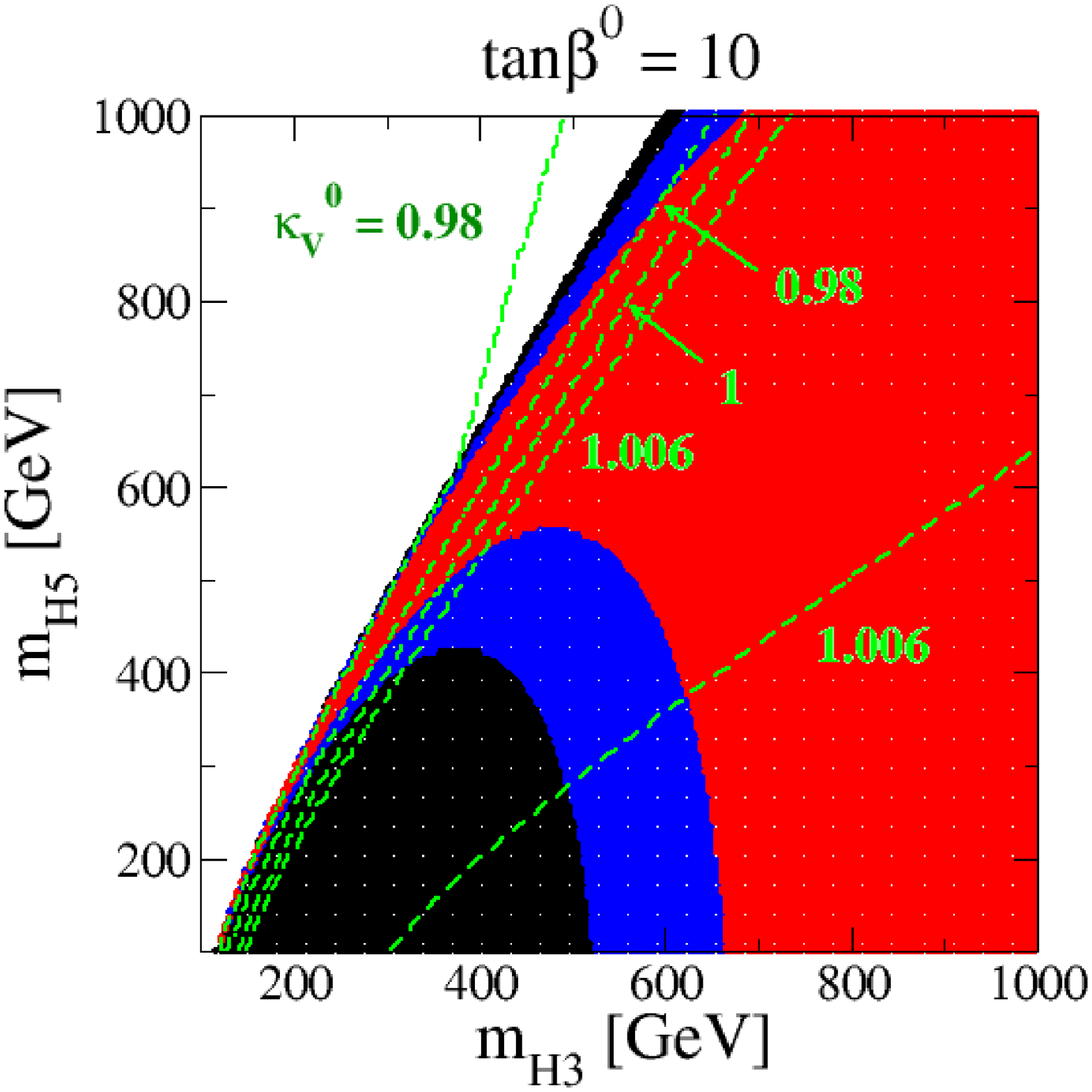}
\caption{The shaded region is allowed by the triviality and the vacuum stability bounds with the required cutoff scale to be larger than $10^{15}$ GeV (red), 
$10^{8}$ GeV (blue) and $10^{4}$ GeV (black) GeV. 
The value of $\tan\beta^0$ is chosen to be $5$ (left panel) and 10 (right panel). 
We take $\rho_1^0=\rho_2^0=\sigma_1^0=\sigma_2^0=0$. 
The green dashed lines show the contour of the $\kappa_V^{0}$ value.}
\label{fig:const2}
\end{center}
\end{figure}

Let us show the previously derived bounds in terms of the masses of extra Higgs bosons, namely, 
the custodial 3-plet mass $m_{H_3}^{}$ and the 5-plet mass $m_{H_5}^{}$ at $\mu_0$. 
In Fig.~\ref{fig:const2}, we show the allowed parameter space on the $m_{H_3}^{}$--$m_{H_5}^{}$ plane with $\rho_1^0 =\rho_2^0 =\sigma_1^0 =\sigma_2^0 =0$ 
and fixed values of $\tan\beta^0$, i.e., $\tan\beta^0 = 5$ (left) and $\tan\beta^0 = 10$ (right). 
Again, we show the contour of the $\kappa_V^{0}$ value by the green dashed curves. 
Similarly to Fig.~\ref{fig:const1}, the black, blue and red shaded regions are allowed by requiring $\Lambda_{\text{cutoff}}$ to be larger than $10^4$, 
$10^8$ and $10^{15}$ GeV, respectively.  
In this plot, the values of $\mu_1^0$ and $\mu_3^0$ are determined for each  point on this plane through Eqs.~(\ref{m5sq}) and (\ref{m3sq}). 
As a typical behavior, larger $m_{H_3}^{}$ and $m_{H_5}^{}$ are allowed with higher $\Lambda_{\text{cutoff}}$ for the case with larger values of $\tan\beta^0$. 
This property can also be seen in Fig.~\ref{fig:const1}, where a larger value of $\mu_1^0$ which provides larger values of $m_{H_3}^{}$ and $m_{H_5}^{}$,  
is allowed for a larger value of $\tan\beta^0$. 
It is also seen that the region with $m_{H_3}^{} \geq m_{H_5}^{}$ and $\kappa_V^{0}>1$ is favored by the triviality and the vacuum stability bounds.

\begin{figure}[!t]
\begin{center}
\includegraphics[width=90mm]{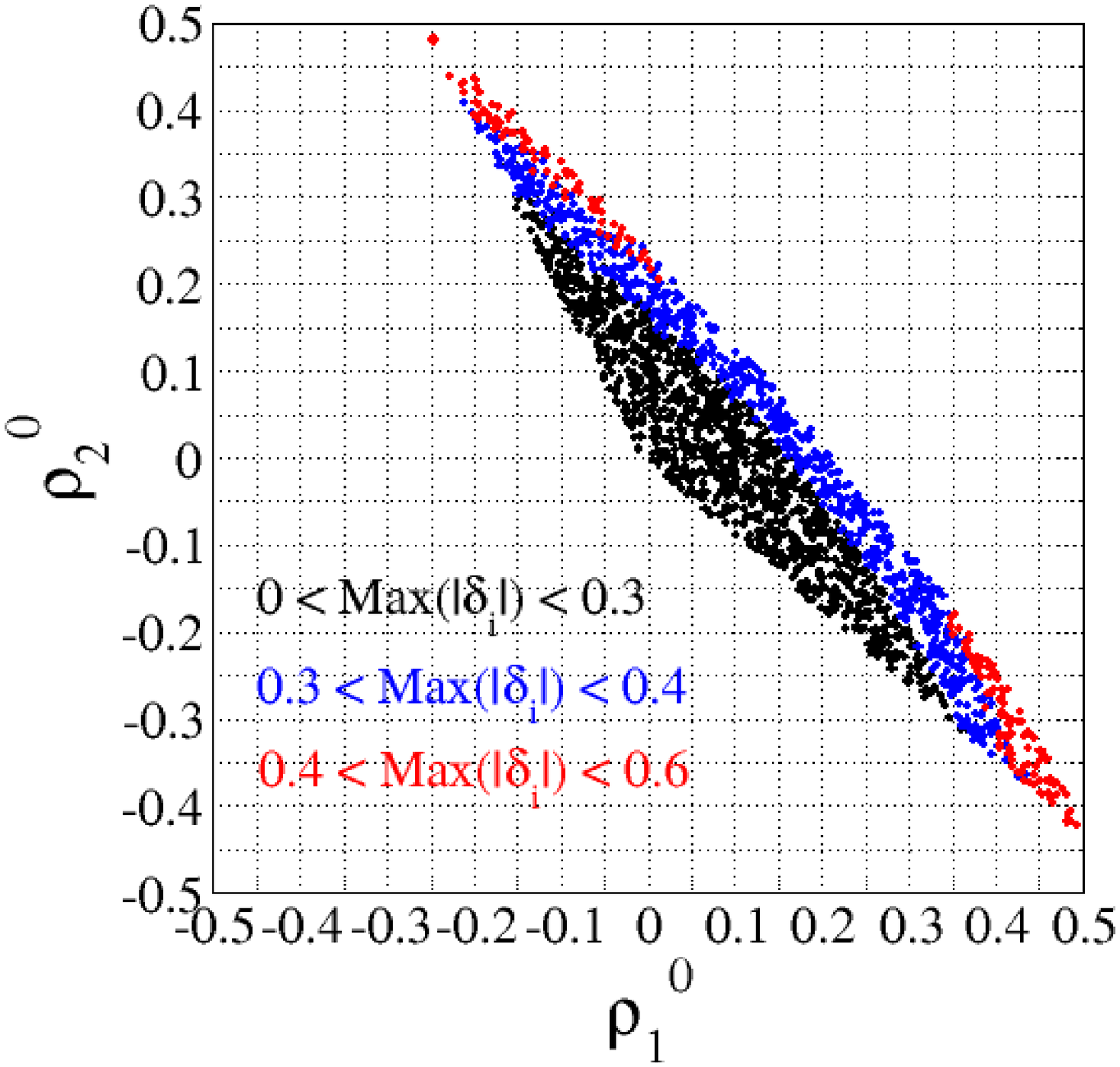}\hspace{-20mm}
\includegraphics[width=90mm]{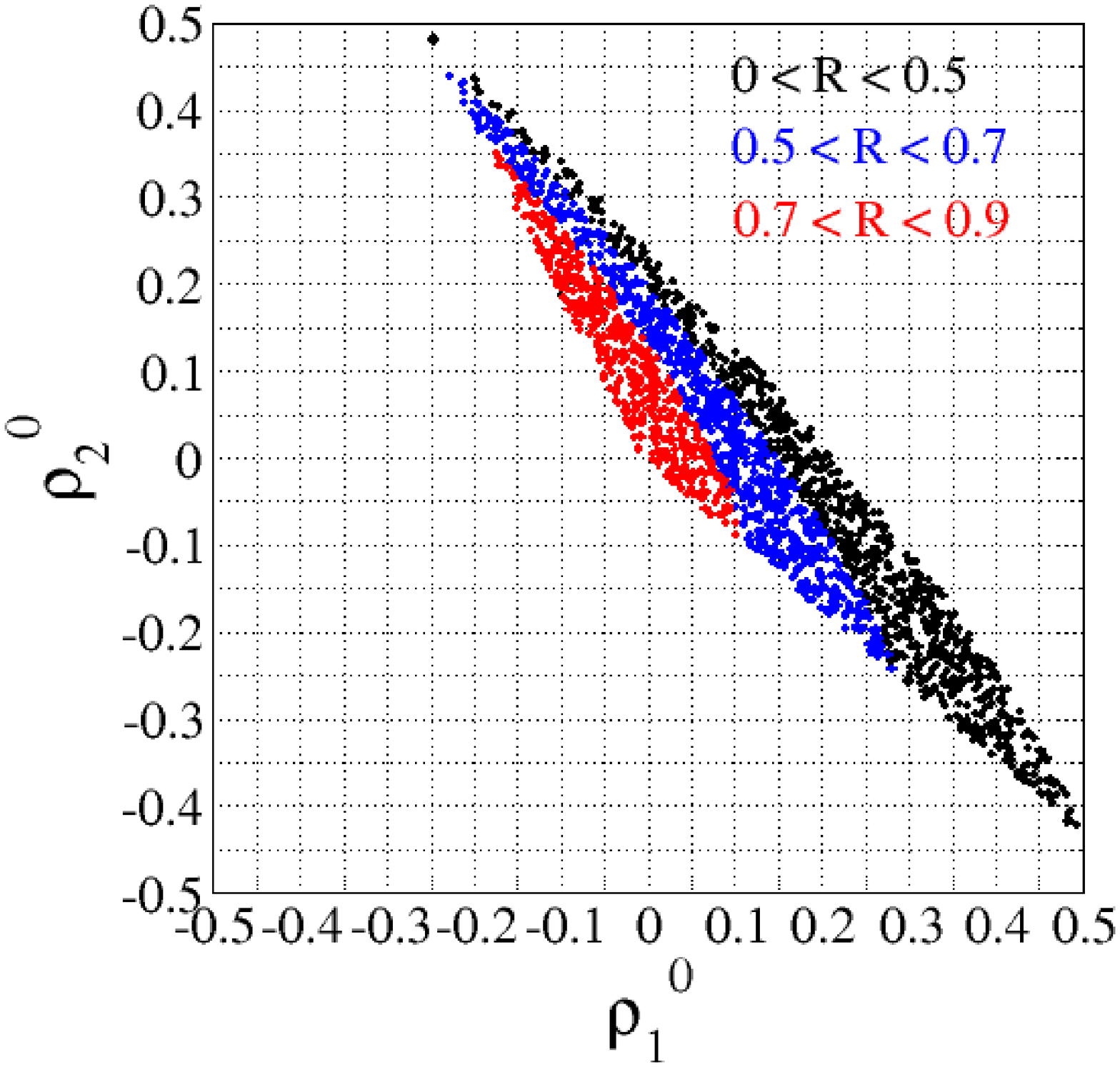}\\ 
\includegraphics[width=90mm]{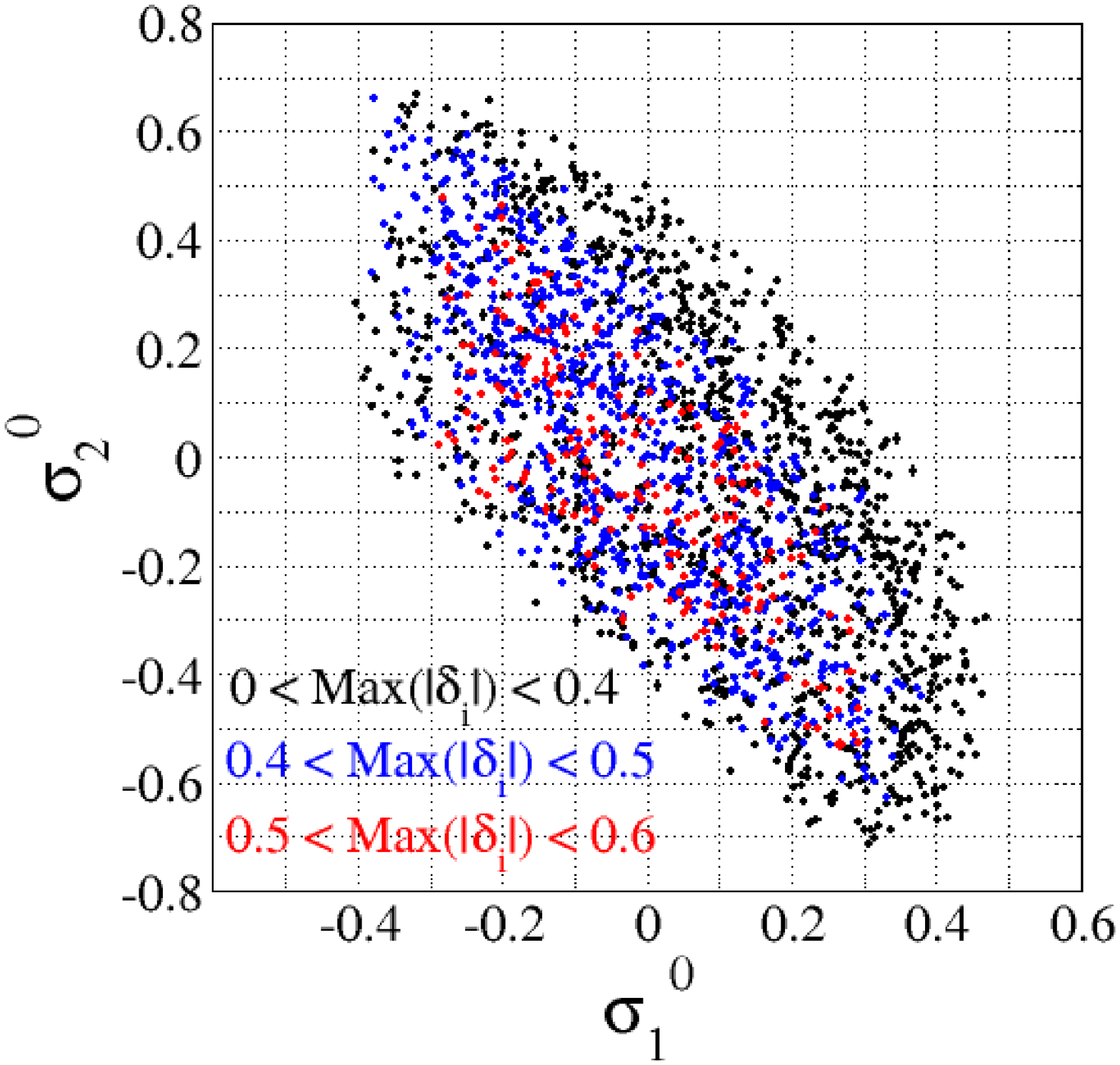}\hspace{-20mm}
\includegraphics[width=90mm]{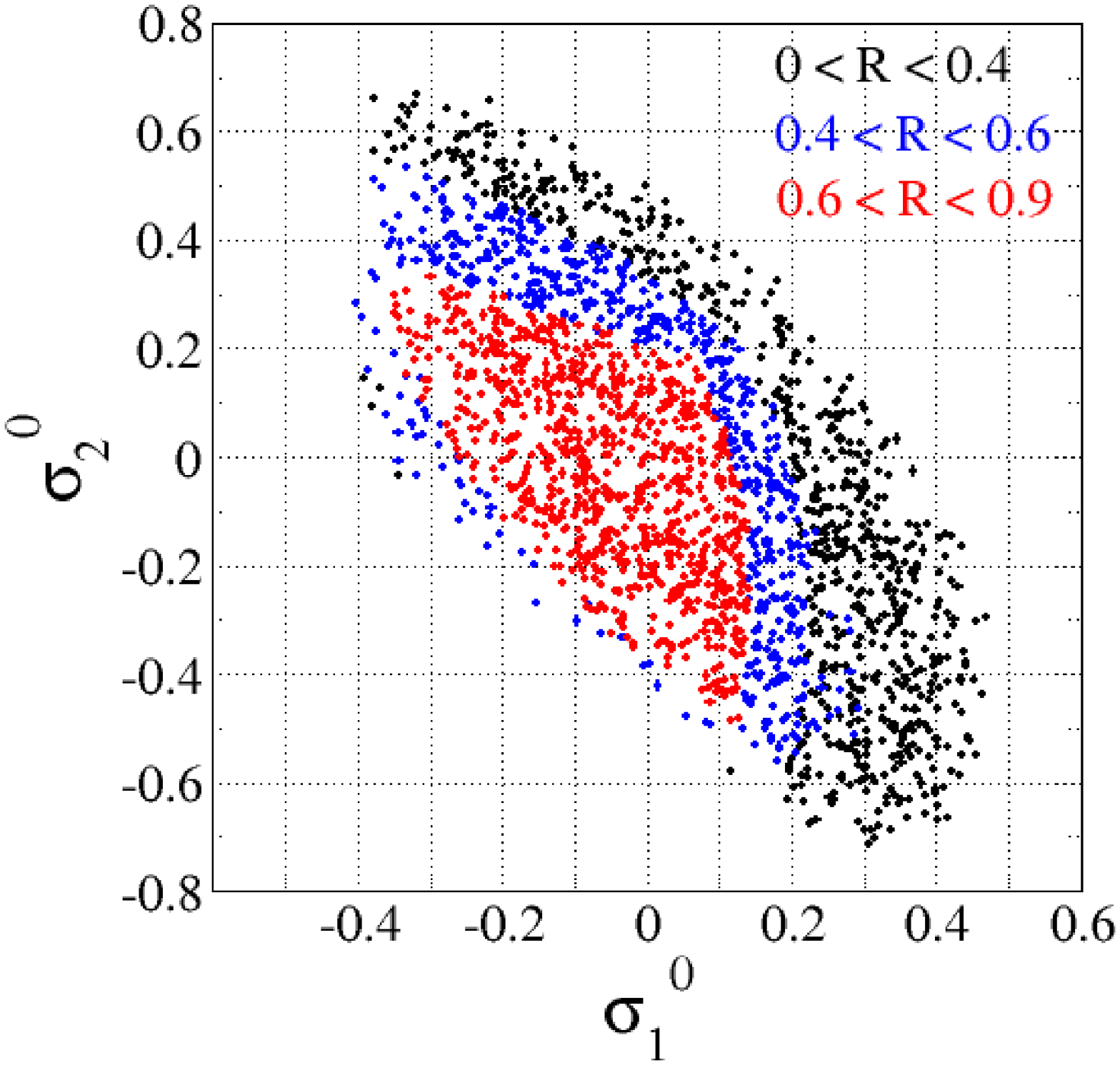}
\caption{
Values of $\text{Max}(|\delta_i|)$ ($i=3,\dots ,7$) (left) 
and $R \equiv \text{Max}(|\delta_i|)/\text{Max}(|\lambda|,|\rho_j|,|\sigma_k|)$ ($j=1,2$ and $k=1,2$) (right) at $\mu= 10^{14}$ GeV on the $\rho_1^0$--$\rho_2^0$ plane (upper panels) and the $\sigma_1^0$--$\sigma_2^0$ plane (lower panels). 
For all the figures, we take $\mu_1^0 = 100$ GeV, $\mu_3^0 = 0$ and $\tan\beta^0 = 5$. 
}
\label{scan}
\end{center}
\end{figure}

Now, let us consider the case with the boundaly conditions different from $\rho_1^0=\rho_2^0=\sigma_1^0 = \sigma_2^0 =0$. 
In Fig.~\ref{scan}, each dot is allowed by the triviality and vacuum stability bounds with $\Lambda_{\text{cutoff}}\geq 10^{15}$ GeV in the case of 
$\mu_1^0 = 100$ GeV, $\mu_3^0=0$ and $\tan\beta^0 = 5$. 
Here, we scan the four inputs $(\rho_1^0,\rho_2^0,\sigma_1^0,\sigma_2^0)$ within the range from $-1$ to $+1$.
From the upper (lower) panels, we can see the allowed region on the $\rho_1^0$--$\rho_2^0$ ($\sigma_1^0$--$\sigma_2^0$) plane. 
We checked that the shape of the allowed region does not change so much if we change the values of $(\mu_1^0,\mu_3^0,\tan\beta^0)$ 
as long as they are allowed with $\Lambda_{\text{cutoff}}\geq 10^{15}$ GeV as shown in Fig.~\ref{fig:const1}. 
In this figure, the dots in the left panels show the range of $\text{Max}(|\delta_i|)$ with $i=3,\dots,7$ 
and those in the right panels represent the range of the ratio $R$ defined by $R \equiv \text{Max}(|\delta_i|)/\text{Max}(|\lambda|,|\rho_j|,|\sigma_k|)$ with $j=1,2$ and $k=1,2$. 
The three different colors show the different ranges of $\text{Max}(|\delta_i|)$ or $R$, where the range is indicated inside the figure.  
We find that at $\mu=10^{14}$ GeV the value of $\text{Max}(|\delta_i|)$ can go up to $\sim 0.6$ which is $\sim g_2^0$, while 
the value of $R$ is smaller than 1.
In addition, by looking at the upper-left figure, 
the value of $\text{Max}(|\delta_i|)\sim 0.6$ is only reached by a large $|\rho_j^0|$ value such as $\rho_{1,2}\simeq 0.4$ with $\rho_{2,1}\simeq -\rho_{1,2}$, while 
in most of the region with $|\rho_j|\lesssim 0.3$, we have a milder value of $\text{Max}(|\delta_i|)\lesssim 0.3$. 
On the contrary by looking at the upper-right figure, we find that a larger value of $R$ (but still less than 1) is obtained 
for a smaller $|\rho_j^0|$ values. 
If we look at the lower-left figure, it is difficult to see a correlation between the value of $\text{Max}(|\delta_i|)$ and the $\sigma_k^0$ parameters.  
This suggests that the value of $\text{Max}(|\delta_i|)$ is almost determined by $\rho_j^0$ which are blind in this plane. 
The upper and lower right figures show a similar behavior of $R$, i.e., smaller values of $|\sigma_k^0|$ gives a larger value of $R$. 
Summarizing we have checked that, if we vary the initial conditions on $\rho_1^0$, $\rho_2^0$, $\sigma_1^0$ and $\sigma_2^0$ in a natural range, 
the custodial symmetry breaking parameters $\delta_i$ keep values smaller than the other parameters in the potential.  

Finally, we show the predictions for the masses of the Higgs bosons at $\mu= 1$ TeV to see how the running parameters $\delta_i$ 
affect the spectrum.
In order to calculate the Higgs boson masses at $\mu > \mu_0$, 
we need to evaluate not only the running of the dimensionless couplings, but also that of the dimensionful parameters $\mu_{1,2,3}$ and $m_{\phi,\chi,\xi}^2$ (their 
one-loop $\beta$-functions are presented in App.~\ref{sec:rge}). 
At a given scale $\mu$, we need to re-impose the tadpole conditions which give three different values of $v_\phi$, $v_\chi$ and $v_\xi$. 
We find that the difference between $v_\chi$ and $v_\xi$ at the TeV scale is quite small, i.e. ${\cal O}(1)$ GeV level, so that 
the mass formulae given in App.~\ref{sec:mass} give a good enough approximation to derive the spectrum at $\mu=1$ TeV. 

\renewcommand\arraystretch{1.4}
\begin{table}[t]
 \centering
 \begin{tabular}{ ccccc || c  c  c  c }\hline
     $m_{H_5}^{}$ & $m_{H_3}^{}$ & $m_{H}^{}$ & $\tan\beta^0$ & $\kappa_V^{0}$ & $(\bar{m}_{H_5^{\pm\pm}}^{},\bar{m}_{H_5^{\pm}}^{},\bar{m}_{H_5^{0}}^{})$ & $(\bar{m}_{H_3^{\pm}}^{},\bar{m}_{H_3^{0}}^{})$ 
     & $\bar{m}_{H^0}$ & $\sin\bar{\gamma}$  \\ \hline\hline
     400 & 300 & 250 & 5  & 1.00 & (589,~591,~592) & (577,~576) & 570 & $-0.14$  \\ \hline
     300 & 400 & 441 & 5  & 1.03 & (521,~522,~522) & (544,~544) & 555 & $-0.011$  \\ \hline
     600 & 650 & 673 & 10  & 1.01 & (951,~951,~951) & (956,~956) & 959 & $-0.013$  \\ \hline
 \end{tabular}
 \caption{(first column): Initial values of $m_{H_5}^{}$, $m_{H_3}^{}$, $m_{H}^{}$, $\tan\beta^0$ and $\kappa_V^0$. 
For all the three sets, we take $\rho_1^0=\rho_2^0=\sigma_1^0=\sigma_2^0=0$.  
(second column): Running masses for the $SU(2)_V$ 5-plet Higgs bosons $(\bar{m}_{H_5^{\pm\pm}}^{},\bar{m}_{H_5^{\pm}}^{},\bar{m}_{H_5^{0}}^{})$, 
the 3-plet Higgs bosons $(\bar{m}_{H_5^{\pm\pm}}^{},\bar{m}_{H_5^{\pm}}^{},\bar{m}_{H_5^{0}}^{})$, the singlet Higgs boson $\bar{m}_{H^0}$, 
and the mixing angle $\bar{\gamma}$ between $H_3^\pm$ and $H_5^\pm$ at $\mu = 1$ TeV. All the masses are given in GeV unit. }
 \label{table1}
\end{table}

In Tab.~\ref{table1}, we show the running masses of the $SU(2)_V$ 5-plet Higgs bosons $(\bar{m}_{H_5^{\pm\pm}}^{},\bar{m}_{H_5^{\pm}}^{},\bar{m}_{H_5^{0}}^{})$, 
the 3-plet Higgs bosons $(\bar{m}_{H_3^{\pm}}^{},\bar{m}_{H_3^{0}}^{})$ and the singlet Higgs boson $\bar{m}_{H^0}$ at $\mu=1$ TeV for the three different sets 
of the initial values at $\mu_0=m_Z^{}$ written in the first column of the table. 
For the input values at $\mu_0$, we here fix $m_{H_5}^{}$, $m_{H_3}^{}$ and $\tan\beta^0$ instead of inputting $\mu_1^0$, $\mu_3^0$ and $\tan\beta^0$, and 
also take $\rho_1^0=\rho_2^0=\sigma_1^0=\sigma_2^0=0$. All the three sets are allowed by both triviality and vacuum stability bounds with $\Lambda_{\text{cutoff}}> 10^{15}$ GeV. 
We note that other choices with non-zero values of the inputs $\rho_{1,2}^0$ and $\sigma_{1,2}^0$ do not change so much the mass spectrum at 1 TeV 
from the results given in this table as long as we assume $\Lambda_{\text{cutoff}}> 10^{15}$ GeV. 
We can see that the breaking of the mass degeneracy among the 5-plet Higgs bosons and that among the 3-plet Higgs bosons 
is only given to be ${\cal O}(1)$ GeV level. 
In addition, the running mixing angle $\bar{\gamma}$ between $H_3^\pm$ and $H_5^\pm$ is given to be $\sim 0.1$ or smaller. 

From the above results, we conclude that in the TeV region 
the mass spectrum of the Higgs bosons or, equivalently, the Higgs potential with the custodial $SU(2)_V$ symmetry
still provides a good approximation to describe the scenario once the loop effect of the custodial symmetry breaking is taken into account. 

Before closing this section, let us briefly comment on the signatures of the 5-plet and 3-plet Higgs bosons and the current bounds on their masses at collider experiments. 
Concerning the 5-plet Higgs bosons, since they do not couple to fermions at tree level, 
their main decay modes are given by diboson channels, i.e., 
$H_5^{\pm\pm} \to W^\pm W^\pm$, $H_5^\pm \to W^\pm Z$ and $H_5^0\to W^+W^-/ZZ$ (see, e.g.,~\cite{CY}). 
In Ref.~\cite{ww}, the 95\% CL upper limit on the branching ratio ($H_5^{\pm\pm} \to W^\pm W^\pm$) 
times the cross section of the vector boson fusion process ($q\bar{q}' \to q\bar{q}' W^\pm W^\pm \to q\bar{q}' H_5^{\pm\pm}$) has been set using the 8 TeV data at the LHC with 
an integrated luminosity of 19.4 fb$^{-1}$. 
From this analysis, the 95\% CL lower bound on the mass of $H_5^{\pm\pm}$ can be extracted to be 
about 300 GeV  when the triplet VEV $v_\Delta$ is taken to be 25 GeV corresponding to $\tan\beta \simeq 3.3$. 
These bounds become weaker for smaller (larger) value of $v_\Delta^{}$ $(\tan\beta)$\footnote{In Ref.~\cite{Const-WW}, 
the mass bound on doubly-charged Higgs bosons $H^{\pm\pm}$ decaying into $W^\pm W^\pm$ 
was also derived in the Higgs triplet model whose 
Higgs sector is composed of one doublet $(Y=1/2)$ plus one triplet $(Y=1)$ fields. 
From the pair production and the associated production with a singly-charged Higgs boson,  
the lower bound on $m_{H^{\pm\pm}}$ was obtained to be about 84 GeV at 95\% CL using the LHC Run-1 data set. 
A similar bound can be applied to the mass of $H_5^{\pm\pm}$ in the GM model without depending on $v_\Delta^{}$. }. 

In Ref.~\cite{wz}, a search for singly-charged Higgs bosons decaying into the $WZ$ mode via the $W$ and $Z$ boson fusion process 
has been performed by using the 13 TeV data set at the LHC with an integrated luminosity of 15.2 fb$^{-1}$. 
The bound is much weaker than that obtained from the search for the $W^\pm W^\pm$ channel.
In fact, for $v_\Delta \lesssim 35$ GeV ($\tan\beta \lesssim 2.3$), no bound ia taken on the mass of $H_5^\pm$ at 95\% CL. 

Concerning the 3-plet Higgs bosons, their phenomenological properties are quite similar to those of singly-charged Higgs bosons  and a
CP-odd Higgs boson in the Type-I 2-Higgs doublet model (2HDM) in the alignment limit~\cite{2hdm}. 
In our notation, $\tan\beta$ plays the same phenomenological role as that in the Type-I 2HDM, i.e., the Yukawa couplings for $H_3^\pm$ and $H_3^0$
are proportional to $\cot\beta$. 
Therefore, the main decay modes of $H_3^\pm$ and $H_3^0$ are typically $tb$ and $t\bar{t}$, 
respectively, as long as these are kinematically allowed. 
For lighter 3-plet Higgs bosons below the $tb$ and $t\bar{t}$ threshold, 
$H_3^\pm\to \tau\nu$ and $H_3^0 \to b\bar{b}/\tau\tau$ can be dominant, respectively. 
A dedicated study for the phenomenology of the 3-plet Higgs boson have been done in Ref.~\cite{CY}. 

\section{Conclusions \label{sec:con}}

We have discussed the high energy behavior of the GM model, particularly shedding light on the effect of the custodial symmetry breaking 
by using the one-loop RGEs. 
In order to obtain a consistent form of the one-loop $\beta$-functions, 
we start from the most general Higgs potential without the custodial $SU(2)_V$ symmetry, 
which is described by 16 independent parameters in the case of CP-conservation. 
The custodial symmetric version of the potential is obtained by taking all the 7 $\delta_i$ parameters, describing the breaking of the custodial symmetry, to be zero. 

We then numerically derived the evolution with energy of $\delta_i$ under the assumption that they all vanish  at $\mu_0 = m_Z^{}$ as initial condition. 
First, we surveyed the parameter region allowed by the triviality and the vacuum stability constraints as a function of the cutoff scale $\Lambda_{\text{cutoff}}$. 
Requiring the model to be consistent up to a high energy scale, e.g. $\Lambda_{\text{cutoff}} \geq  10^{15}$ GeV, we obtain
a strong correlation between the dimensionful trilinear coupling $\mu_1$ and $\tan\beta$ and between the mass of the custodial 5-plet Higgs boson 
and that of the 3-plet Higgs boson at $\mu=\mu_0$. 
We then extracted the typical size of the $\delta_i$ parameters at high energies. 
We found that, in the configurations with $\Lambda_{\text{cutoff}} \geq 10^{15}$ GeV, 
the maximal value of $|\delta_i|$ can be up to $\sim 0.6$ at $\mu = 10^{14}$ GeV, and it is smaller than the maximal value of the 
input parameters in the potential ($\lambda$, $\rho_{1,2}$ and $\sigma_{1,2}$).  

In addition, in order to quantify the effects of the custodial symmetry breaking, we derived the running masses of the Higgs bosons
and the running mixing angle $\bar{\gamma}$ between the $H_3^\pm$ and $H_5^\pm$ at $\mu = 1$ TeV.  
We found that the deviation from the custodial symmetric limit is quite small, namely, the mass splitting among the Higgs bosons belonging to the same $SU(2)_V$ multiplet  
is of the order of 1 GeV, and $\sin\bar{\gamma}$  $\sim 0.1$. 
This means that once custodial symmetry is realized at low energy ($m_Z^{}$ scale), 
it also approximately holds at the TeV scale which is now being surveyed at the LHC experiments.  

\begin{appendix}

\section{Relations among scalar fields \label{sec:rel} }

Relations between the fields $\Phi$ and $\Delta$ defined in Eq.~(\ref{eq:Higgs_matrices}) and $\phi$, $\chi$ and $\xi$ defined in Eq.~(\ref{par}) are given as 
\begin{align}
\text{tr}(\Phi^\dagger \Phi) &= 2 \phi^\dagger \phi, \\
\text{tr}(\Delta^\dagger \Delta) &= 2\text{tr}(\chi^\dagger \chi) + \text{tr}(\xi^2), \\
\text{tr}\left(\Phi^\dagger \frac{\tau^a}{2}\Phi\frac{\tau^b}{2}\right)(P^\dagger \Delta P)^{ab}&= -\frac{1}{\sqrt{2}}\phi^\dagger \xi \phi 
-\frac{1}{2}[\phi^T (i\tau_2) \chi^\dagger \phi + \text{h.c.}], \\
\text{tr}\left(\Delta^\dagger t^a \Delta t^b\right)(P^\dagger \Delta P)^{ab} &= 6\sqrt{2} \text{tr}(\chi^\dagger\chi\xi), \\
\left[\text{tr}(\Delta^\dagger \Delta) \right]^2 &= 4\left[\text{tr}(\chi^\dagger \chi)\right]^2 + 2\text{tr}(\xi^4) + 4 \text{tr}(\chi^\dagger\chi)\text{tr}(\xi^2), \\
\text{tr}(\Delta^\dagger \Delta\Delta^\dagger \Delta) &= 
6\left[\text{tr}(\chi^\dagger \chi)\right]^2 -4\text{tr}(\chi^\dagger \chi \chi^\dagger \chi) +2\text{tr}(\xi^4)  + 4 \text{tr}(\chi^\dagger\xi)\text{tr}(\xi \chi), \\ 
\text{tr}\left(\Phi^\dagger\frac{\tau^a}{2}\Phi\frac{\tau^b}{2}\right)
\text{tr}(\Delta^\dagger t^a\Delta t^b)
&= -\phi^\dagger \phi \text{tr}(\chi^\dagger \chi) + 2\text{tr}\phi^\dagger \chi \chi^\dagger \phi +\sqrt{2}(\phi^\dagger \chi\xi \phi^c + \text{h.c.}). 
\end{align}
We note $\text{tr}(\xi^4)=[\text{tr}(\xi^2)]^2/2$.

\section{Mass Formulae \label{sec:mass}}

Let us present the mass formulae for the Higgs bosons of the GM model with the general potential defined in Eq.~(\ref{pot_gen}) and  $v_\chi=v_\xi = v_\Delta^{}$. 

The mass of the doubly-charged scalar states $\chi^{\pm\pm} (\equiv {\cal H}_5^{  \pm\pm})$ is given by 
\begin{align}
m_{{\cal H}_5^{\pm\pm }}^2 = \frac{v}{4}\left[4\sqrt{2}s_\beta t_\beta \mu_2 -2 c_\beta \mu_3 - v(c_\beta^2\rho_2 +2s_\beta^2\sigma_2 +\sqrt{2}s_\beta^2\sigma_4) \right]. 
\end{align}

For the singly-charged scalar states, 
the weak eigenstates ($\xi^\pm$, $\phi^\pm$, $\chi^\pm$) are related to the mass eigenstates 
($G^\pm$, ${\cal H}_3^\pm$, ${\cal H}_5^\pm$), with $G^\pm$ being the Nambu-Goldstone (NG) bosons to be absorbed into the longitudinal components of the $W^\pm$ bosons, 
by the following orthogonal transformation:
\begin{align}
\begin{pmatrix}
\xi^\pm \\
\phi^\pm \\
\chi^\pm 
\end{pmatrix} = 
\begin{pmatrix}
\frac{1}{\sqrt{2}} &0& -\frac{1}{\sqrt{2}} \\
0&1&0 \\
\frac{1}{\sqrt{2}} &0& \frac{1}{\sqrt{2}} 
\end{pmatrix}
\begin{pmatrix}
c_\beta & s_\beta & 0\\
s_\beta & -c_\beta & 0\\
0&0&1
\end{pmatrix} 
\begin{pmatrix}
1 & 0 & 0\\
0 & c_\gamma & -s_\gamma \\
0 & s_\gamma & c_\gamma \\
\end{pmatrix} 
\begin{pmatrix}
G^\pm \\
{\cal H}_3^{\pm}\\
{\cal H}_5^{\pm}
\end{pmatrix}. \label{sing}
\end{align}
The mixing angle $\gamma$ and the mass eigenvalues 
$m_{{\cal H}_3^\pm}^2$ and $m_{{\cal H}_5^\pm}^2$ for the ${\cal H}_3^{\pm}$ and ${\cal H}_5^{\pm}$ states, respectively, are expressed by 
\begin{align}
m_{{\cal H}_3^\pm}^2 & = (M_\pm^2)_{11} c_{\gamma}^2 + (M_\pm^2)_{22} s_{\gamma}^2  + 2(M_\pm^2)_{12} c_{\gamma}s_{\gamma}, \\
m_{{\cal H}_5^\pm}^2 & = (M_\pm^2)_{11} s_{\gamma}^2 + (M_\pm^2)_{22} c_{\gamma}^2  - 2(M_\pm^2)_{12} c_{\gamma}s_{\gamma}, \\
\tan 2\gamma  &= \frac{2(M_\pm^2)_{12}}{(M_\pm^2)_{11}-(M_\pm^2)_{22}}, 
\end{align}
where
\begin{align}
(M_\pm^2)_{11} &= \frac{v}{8}\left[\frac{4}{c_\beta}(\mu_1 + \sqrt{2}\mu_2) -v\left(\sigma_2 + \sqrt{2}\sigma_4 \right)\right], \\
(M_\pm^2)_{22} &= \frac{v}{8}\left[4s_\beta t_\beta (\mu_1 + \sqrt{2}\mu_2) -4c_\beta \mu_3 
-v\left(s_\beta^2 \sigma_2 +5\sqrt{2}s_\beta^2\sigma_4 -2c_\beta^2\rho_5 \right)
\right], \\
(M_\pm^2)_{12} &= \frac{v}{8}\left[-4t_\beta(\mu_1 - \sqrt{2} \mu_2) -vs_\beta(\sigma_2 - \sqrt{2}\sigma_4)\right]. 
\end{align}

For the CP-odd scalar states, 
the weak eigenstates ($\chi_i$, $\phi_i$) are related to the mass eigenstates 
($G^0$, ${\cal H}_3^0$), with $G^0$ being the NG boson to be absorbed into the longitudinal component of the $Z$ boson, 
by the following orthogonal transformation:
\begin{align}
\begin{pmatrix}
\chi_i \\
\phi_i 
\end{pmatrix} = 
\begin{pmatrix}
c_\beta  & -s_\beta \\
s_\beta & c_\beta
\end{pmatrix} 
\begin{pmatrix}
G^0 \\
{\cal H}_3^0\\
\end{pmatrix}. 
\end{align}
The squared mass $m_{{\cal H}_3^{0  }}^2$ for ${\cal H}_3^{ 0}$ is expressed by 
\begin{align}
m_{{\cal H}_3^{0  }}^2 = \frac{\sqrt{2}\mu_2 v}{c_\beta} - \frac{\sqrt{2}}{4}v^2\sigma_4. 
\end{align}

Finally, for the CP-even Higgs states, we define the following basis:
\begin{align}
\begin{pmatrix}
\xi_r\\
\phi_r\\
\chi_r
\end{pmatrix}
 = \left(
\begin{array}{ccc}
\frac{1}{\sqrt{3}} &0& -\sqrt{\frac{2}{3}}\\
0 & 1 &0\\
\sqrt{\frac{2}{3}} & 0 & \frac{1}{\sqrt{3}}
\end{array}\right)
\begin{pmatrix}
\tilde{H}\\
\tilde{h} \\
\tilde{\cal H}_5^{0}
\end{pmatrix}, 
\end{align}
where the three states $\tilde{H}$, $\tilde{h}$ and $\tilde{\cal H}_5^0$ are not mass eigenstates in general. 
The squared mass matrix elements, in the basis $(\tilde{H}$, $\tilde{h}$ and $\tilde{\cal H}_5^{0})$, are expressed as
\begin{align}
&(M_{\text{even}}^2)_{11} = \frac{v}{6}\left[2s_\beta t_\beta (\mu_1 +2\sqrt{2}\mu_2) +\frac{3}{2}c_\beta \mu_3 + vc_\beta^2(2\rho_1+2\rho_2+\rho_3+2\rho_4)  \right],\label{even11} \\
&(M_{\text{even}}^2)_{22}  = 2s_\beta^2 v^2 \lambda, \label{even22}\\
&(M_{\text{even}}^2)_{12}  = \frac{vs_\beta}{\sqrt{6}}\left[-\mu_1 -2\sqrt2{}\mu_2 + vc_\beta(\sigma_1+\sigma_2+\sigma_3+\sqrt{2}\sigma_4) \right], \label{even12}\\
&(M_{\text{even}}^2)_{13}  = \frac{v}{6\sqrt{2}}\left[-4s_\beta t_\beta(\mu_1 - \sqrt{2}\mu_2) +vc_\beta^2(2\rho_1+2\rho_2-2\rho_3-\rho_4)\right], \\
&(M_{\text{even}}^2)_{23}  = \frac{vs_\beta}{2\sqrt{6}}\left[2\sqrt{2}\mu_1-4\mu_2 + vc_\beta (\sqrt{2}\sigma_1 + \sqrt{2}\sigma_2 -2\sqrt{2}\sigma_3-\sigma_4) \right], \\
&\hspace{-0.5cm}(M_{\text{even}}^2)_{33}  = \frac{v}{6}\left[
2\sqrt{2}s_\beta t_\beta (\sqrt{2}\mu_1 + \mu_2) -3c_\beta \mu_3 +vc_\beta^2(\rho_1+\rho_2+2\rho_3-2\rho_4)-\frac{9}{\sqrt{2}}vs_\beta^2\sigma_4\right]. 
\end{align}
The relation of the basis $(\tilde{H},\tilde{h},\tilde{H}_5^0)$ to the mass eigenstates is obtained by an orthogonal transformation: 
\begin{align}
\begin{pmatrix}
\tilde{H}\\
\tilde{h} \\
\tilde{\cal H}_5^{0}
\end{pmatrix}
 = R_{\text{even}}
\begin{pmatrix}
H \\
h \\
{\cal H}_5^{0}
\end{pmatrix}, 
\end{align} 
where $R_{\text{even}}$ can be expressed in terms of three independent mixing angles. 

In the custodial symmetric limit defined in Eq.~(\ref{custodial}), we obtain 
\begin{align}
&(M_{\pm}^2)_{12} = (M_{\text{even}}^2)_{13} = (M_{\text{even}}^2)_{23} = 0,  \label{mix2} \\
&(M_{\pm}^2)_{22}\, (=m_{{\cal H}_5^{\pm}}^2) = (M_{\text{even}}^2)_{33}\, (=m_{{\cal H}_5^0}^2) = m_{{\cal H}_5^{\pm\pm}}^2, \\
&(M_{\pm}^2)_{11}\, (=m_{{\cal H}_3^{\pm}}^2) = m_{{\cal H}_3^0}^2. 
\end{align}
Therefore, we can clearly reproduce the custodial symmetric results, namely, 
$({\cal H}_5^{\pm\pm},{\cal H}_5^{\pm},{\cal H}_5^0)$ and $({\cal H}_3^{\pm},{\cal H}_3^0)$ 
are the custodial 5-plet ($H_5^{\pm\pm},H_5^{\pm},H_5^0$) and the 3-plet $(H_3^{\pm},H_3^0)$, respectively. 
Because of the no mixing displayed in Eq.~(\ref{mix2}), the Higgs bosons belonging to the different custodial multiplets are not mixed with each other. 
In addition, the degeneracy of masses for Higgs bosons belonging to the same custodial multiplet follows: 
\begin{align}
m_{H_5^{}}^2 &= \frac{v}{4}\left[4s_\beta t_\beta \mu_1 - 2c_\beta\mu_3  -v(c_\beta^2\rho_2 + 3s_\beta^2\sigma_2)\right], \label{m5sq}\\
m_{H_3^{}}^2 &= \frac{v}{c_\beta}\mu_1 -\frac{v^2}{4}\sigma_2.  \label{m3sq}
\end{align}
For the CP-even Higgs bosons, the $3\times 3$ matrix $R_{\text{even}}$ becomes the block diagonal form as $R_{\text{even}} = \text{diag}(R(\alpha),1)$ which 
is described by only one mixing angle $\alpha$. 
We thus express the custodial singlet Higgs bosons $H$ and $h$ by the linear combination of the $\tilde{H}$ and $\tilde{h}$ states as:
\begin{align}
\begin{pmatrix}
\tilde{H}\\
\tilde{h}\\
\end{pmatrix} = R(\alpha)
\begin{pmatrix}
H\\
h
\end{pmatrix}. 
\end{align} 
The two squared mass eigenvalues and the mixing angle $\alpha$ are expressed as 
\begin{align}
m_{H}^2 & = (M_{\text{even}}^2)_{11} c_{\alpha}^2 + (M_{\text{even}}^2)_{22} s_{\alpha}^2  + 2(M_{\text{even}}^2)_{12} c_{\alpha}s_{\alpha}, \\
m_h^2   & = (M_{\text{even}}^2)_{11} s_{\alpha}^2 + (M_{\text{even}}^2)_{22} c_{\alpha}^2  - 2(M_{\text{even}}^2)_{12} c_{\alpha}s_{\alpha}, \\
\tan 2\alpha  &= \frac{2(M_{\text{even}}^2)_{12}}{(M_{\text{even}}^2)_{11}-(M_{\text{even}}^2)_{22}},  \label{tan2a}
\end{align}
where 
\begin{align}
(M_{\text{even}}^2)_{11} &= \frac{v}{8}\left[ 8s_\beta t_\beta \mu_1 + 2c_\beta\mu_3 +vc_\beta^2(6 \rho_1 + 7\rho_2)  \right], \\
(M_{\text{even}}^2)_{22} & = 2v^2s_\beta^2 \lambda, \\
(M_{\text{even}}^2)_{12} & = \frac{\sqrt{6}}{8}vs_\beta\left[-4\mu_1 + vc_\beta(2\sigma_1+3\sigma_2) \right]. 
\end{align}

\section{$\beta$-functions \label{sec:rge}}

In this Appendix, we give the analytic expressions of the one-loop $\beta$-functions for all the model parameters. 
The definition of the $\beta$-function is given in Eq.~(\ref{betaf}). 

The $\beta$-functions for the 3 gauge couplings $g_i$ ($i=1,2,3$) and the Yukawa couplings for the top ($y_t$) and bottom ($y_b$) quarks are given by  
\begin{align}
\beta(g_3)&=\frac{g_3^3}{16\pi^2}(-7), \quad \beta(g_2)=\frac{g_2^3}{16\pi^2}\left(-\frac{11}{6}\right), \quad \beta(g_1)=\frac{g_1^3}{16\pi^2}\frac{47}{6}, \\
\beta(y_t)&=\frac{1}{16\pi^2}\left[\frac{9}{2}y_t^3+\frac{3}{2}y_b^3 -y_t\left(8g_3^2+\frac{9}{4}g_2^2+\frac{17}{12}g_1^2\right)\right], \\
\beta(y_b)&=\frac{1}{16\pi^2}\left[\frac{9}{2}y_b^3+\frac{3}{2}y_t^3 -y_b\left(8g_3^2+\frac{9}{4}g_2^2+\frac{5}{12}g_1^2\right)\right]. 
\end{align}

For the 10 dimensionless parameters in the potential given in Eq.~(\ref{pot_gen}), we have 
\begin{align}
16\pi^2\beta(\lambda) & =  \frac{3}{8} \left(3 g_2^4+2g_2^2g_1^2+g_1^4 \right)   +24\lambda^2-6 (y_t^4+y_b^4)
+3 \sigma_1^2+3 \sigma_1 \sigma_2+\frac{5 \sigma_2^2}{4}+6 \sigma_3^2+2 \sigma_4^2 \notag\\
&-3\lambda(g_1^2  +3 g_2^2-4y_t^2-4y_b^2), \\
16\pi^2\beta(\rho_1)& = 15 g_2^4-12 g_1^2 g_2^2+6 g_1^4
+28 \rho_1^2+24 \rho_1 \rho_2+6 \rho_2^2+6 \rho_4^2+4 \rho_4 \rho_5+3\rho_5^2\notag\\
&+2 \sigma_1^2+2 \sigma_1 \sigma_2-12\rho_1( g_1^2 + 2g_2^2), \\
16\pi^2\beta(\rho_2)& =  24 g_1^2 g_2^2-6 g_2^4+24\rho_1 \rho_2+18 \rho_2^2-2 \rho_5^2+\sigma_2^2-12 \rho_2 (2g_2^2 + g_1^2), \\
16\pi^2\beta( \rho_3)& =  2 \left(3 g_2^4+22 \rho_3^2+3 \rho_4^2+2\rho_4 \rho_5+\rho_5^2+2 \sigma_3^2-12 g_2^2 \rho_3\right), \\
16\pi^2\beta( \rho_4)& =  2 \Big[3 g_2^4+\rho_4 \left(8 \rho_1+6 \rho_2+10 \rho_3+4\rho_4 \right)+2 \rho_5 (\rho_1+\rho_2+\rho_3)\notag\\
&+\rho_5^2+2 \sigma_1 \sigma_3+\sigma_2 \sigma_3+\sigma_4^2-3\rho_4( g_1^2 +4 g_2^2)\Big], \\
16\pi^2\beta(\rho_5)& =  2 \Big[3 g_2^4+\rho_5 \left(2 \rho_1+4 \rho_3+8 \rho_4+5\rho_5\right)-\sigma_4^2 -3\rho_5(4 g_2^2 +g_1^2)\Big], \\
16\pi^2\beta(\sigma_1)& =  3 g_1^4-6g_1^2g_2^2+6 g_2^4 +2 \sigma_1 \left(6 \lambda +8 \rho_1+6 \rho_2+2\sigma_1\right) 
+2 \sigma_2 (2 \lambda +3 \rho_1+\rho_2) \notag\\
&+2\left(6 \rho_4 \sigma_3+2\rho_5 \sigma_3+\sigma_4^2\right)+\sigma_2^2
-\frac{3}{2}\sigma_1(5g_1^2 +11g_2^2-4y_t^2-4y_b^2), \\
16\pi^2\beta(\sigma_2)& = 12g_1^2g_2^2
+4\sigma_2 [\lambda +\rho_1+2 (\rho_2+\sigma_1)+\sigma_2] +4 \sigma_4^2\notag\\
& -\frac{3}{2}\sigma_2(5g_1^2+11g_2^2 -4 y_t^2-4y_b^2),\\
16\pi^2\beta(\sigma_3)& = 3 g_2^4 + 2\sigma_3\left(6 \lambda +10 \rho_3+4 \sigma_3\right) + (3 \rho_4+\rho_5) (2\sigma_1+\sigma_2)+4 \sigma_4^2\notag\\
& -\frac{3}{2}\sigma_3( g_1^2 +11 g_2^2-4y_t^2-4y_b^2), \\
16\pi^2\beta(\sigma_4)& =  \frac{\sigma_4}{2}  \left[4 \left(2 \lambda +2\rho_4-\rho_5
+ 2\sigma_1+2\sigma_2+4\sigma_3\right)-3(3g_1^2+11 g_2^2-4y_t^2-4y_b^2)\right]. 
\end{align}

Finally, the $\beta$-functions for the dimensionful trilinear ($\mu_{1,2,3}$) and bilinear ($m^2_{\phi,\chi,\xi}$) couplings are given by
\begin{align}
16\pi^2\beta(\mu_1)&= \frac{\mu_1}{2}(8\lambda +16 \sigma_3 -3g_1^2 -21g_2^2 +12y_t^2 + 12y_b^2)+16\mu_2\sigma_4 -2\mu_3\sigma_2, \\
16\pi^2\beta(\mu_2)&=4\mu_1\sigma_4+\frac{\mu_2}{2}\left(8\lambda + 8\sigma_1 + 12\sigma_2 -9g_1^2 -21g_2^2 +12y_t^2+12y_b^2  \right)-2\mu_3\sigma_4, \\
16\pi^2\beta(\mu_3)&=-2\mu_1\sigma_2 -8\mu_2\sigma_4 + 2\mu_3(2\rho_1+4\rho_2+4\rho_4-2\rho_5-3g_1^2-9g_2^2), \\
16\pi^2\beta(m_\phi^2)&=\frac{3}{2}m_\phi^2(8\lambda -g_1^2 - 3g_2^2 +4y_t^2 +4y_b^2) +3m_\chi^2(2\sigma_1 + \sigma_2)+12m_\xi^2\sigma_3 \notag\\
& + 3\mu_1^2 +12\mu_2^2, \\
16\pi^2\beta(m_\chi^2)&= 2m_\phi^2(2\sigma_1 + \sigma_2)+2m_\chi^2(8\rho_1 +6 \rho_2 - 3g_1^2 - 6g_2^2)+4m_\xi^2(3\rho_4 +\rho_5)\notag\\
&+4\mu_2^2 + 2\mu_3^2, \\
16\pi^2\beta(m_\xi^2)&= 4m_\phi^2\sigma_3 +2m_\chi^2(3\rho_4+\rho_5) +4m_\xi^2\left(5\rho_3 - 3g_2^2 \right)+\mu_1^2+\mu_3^2. 
\end{align}

\end{appendix}

\end{document}